\documentclass[twocolumn]{aastex631}
\usepackage{graphicx}
\usepackage{mathtools}
\graphicspath{{./figures/}}
\usepackage{amsmath}
\usepackage[utf8]{inputenc}
\usepackage[OT6,T1]{fontenc}
\usepackage{float}

\newcommand{\Msolar}{M$_{\odot}$}

\newcommand{\kms}{km s$^{-1}$}
\newcommand{\vt}{$v_{t}$}
\newcommand{\logg}{log$_{10}\left(g\right)$}
\newcommand{\teff}{$T_\mathrm{eff}$}
\newcommand{\vsini}{$v\,$sin$\,i$}
\newcommand{\halpha}{H$\alpha$}

\defcitealias{Milliman2015}{M15}
\defcitealias{Tautv2000}{T00}
\defcitealias{Pancino2010}{P10}
\defcitealias{Jacobson2011}{J11}
\defcitealias{Overbeek2015}{O15}
\defcitealias{Tautv2005}{T05}
\defcitealias{Yong2005}{Y05}
\defcitealias{Liu2019}{L19}

\shorttitle{Barium Abundances of Blue Straggler Stars in NGC 7789 and M67}
\shortauthors{Nine et al.}

\begin{document}

\title{WIYN Open Cluster Study. XC. Barium Surface Abundances of Blue Straggler Stars in the Open Clusters NGC 7789 and M67}

\author[0000-0002-6478-0611]{Andrew C. Nine}
\affiliation{Department of Astronomy, University of Wisconsin-Madison, 475 N Charter St., Madison, WI 53706, USA}
\email{anine@astro.wisc.edu}

\author[0000-0002-7130-2757]{Robert D. Mathieu}
\affiliation{Department of Astronomy, University of Wisconsin-Madison, 475 N Charter St., Madison, WI 53706, USA}

\author[0000-0001-7203-8014]{Simon C. Schuler}
\affiliation{Department of Physics and Astronomy, University of Tampa, 401 W Kennedy Blvd., Tampa, FL 33606, USA}

\author[0000-0002-8308-1998]{Katelyn E. Milliman}
\affiliation{Department of Space Studies, American Public University, 111 W Congress St., Charles Town, WV 25414, USA}

\begin{abstract}

We investigate barium (Ba) abundances in blue straggler stars (BSSs) in two open clusters, NGC 7789 (1.6 Gyr) and M67 (4 Gyr), as signatures of asymptotic-giant-branch (AGB) mass transfer. We combine our findings with previous Ba abundance analyses in NGC 6819 (2.5 Gyr) and NGC 188 (7 Gyr). Out of 35 BSSs studied in NGC 7789, NGC 6819, and M67, 15 (43$\pm$11\%) are Ba-enriched; no BSSs in NGC 188 are Ba-enriched. The Ba abundances of enriched BSSs show an anticorrelation with cluster age, ranging from an enrichment of [Ba/Fe]~$\sim$~+1.5 dex in NGC 7789 to [Ba/Fe]~$\sim$~+1.0 dex in M67. The Ba-enriched BSSs all lie in the same region of the HR diagram, irrespective of cluster age or distance from the main-sequence turnoff. Our data suggest a link between AGB donor mass and mass-transfer efficiency in BSSs, in that less massive AGB donors tend to undergo more conservative mass transfer. We find that 40$\pm$16\% of the Ba-enriched BSSs are in longer-period spectroscopic binaries with orbital periods less than 5000 days. Those Ba-enriched BSSs that do not exhibit radial-velocity variability suggest AGB mass-transfer in wide binaries by either wind mass transfer or wind Roche-lobe overflow. Given the preponderance of long orbital periods in the BSSs of M67 and NGC 188 and the frequency of Ba enrichment in NGC 7789, NGC 6819, and M67, it may be that AGB mass transfer is the dominant mechanism of BSS formation in open clusters older than 1 Gyr.
    \vspace{1cm}
\end{abstract}

\section{Introduction}

Interactions within and between binary stars in open clusters can produce stars that differ from the standard picture of single-star  evolution. The most well-known of these alternative stellar evolution products are the blue straggler stars (BSSs). First observed in the globular cluster M3 by \cite{Sandage1953} and given their present name by \cite{Burbidge1958}, the BSSs are stars that are bluer or brighter than the main-sequence (MS) turnoff (TO) in a cluster color-magnitude diagram (CMD). The BSSs are thought to form through three pathways: the merger of stars either through wind-driven angular momentum loss or through Kozai-Lidov oscillations (\citealt{Andronov2006, Perets2009}); the collision of two stars in a dynamical encounter (\citealt{Hills1976, PZ2010}); or mass transfer from an evolved star onto a MS companion (\citealt{McCrea1964, Chen2008, Mathieu2009}). 

Subsequent observations of open clusters have revealed the presence of alternative stellar evolution products across the CMD beyond BSSs, including the yellow straggler stars (\citealt{Strom1971}), which lie between the terminal-age MS and the red giant branch and are thought to be evolved BSSs; the sub-subgiants (\citealt{Mathieu2003}), which are fainter than the subgiant branches of their host clusters; the red straggler stars (\citealt{Geller2017}), which lie to the red of the red giant branches; and the blue lurkers (\citealt{Leiner2019}), which are anomalously rapidly rotating stars on the MS that are either in long-period spectroscopic binaries ($P_{\mathrm{orb}}>100$~days) or are not detected to be velocity variable, and are thought to be mass-transfer products that have not gained sufficient mass to appear as classical BSSs (\citealt{Nine2023}). More than 25\% of all evolved stars in older open clusters are the products of binary stellar evolution (\citealt{MathieuLeiner2019ASEP}).

Previous observations of the older open clusters M67 (4 Gyr, corresponding to a turnoff mass of 1.3~\Msolar; \citealt{BN2007}) and NGC 188 (7 Gyr, 1.1\Msolar; \citealt{Sarajedini1999}) have found that most BSSs in these clusters are single-lined spectroscopic binaries with orbital periods of order 1000 days (\citealt{Latham2007, Geller2011}), indicating that mass transfer is the predominant mechanism for the creation of open cluster BSSs. Such long orbital periods are indicative of Case C mass transfer from an asymptotic-giant-branch (AGB) donor, which would leave behind the degenerate core as a carbon-oxygen (CO) white dwarf (WD; \citealt{Kippenhahn1967, Lauterborn1970, Paczynski1971}). The secondary mass distribution of the long-period binary BSSs in NGC 188 shows a sharp peak at 0.5 \Msolar\, (\citealt{Geller2011}), and ultraviolet photometric observations of the BSSs in NGC 188 conducted with the Hubble Space Telescope (HST) detected four hot ($>13,000$~K) WD companions and provided evidence for the presence of three cooler ($>10,000$~K) WD companions (\citealt{Gosnell2015}). Overall, WD companions of these temperatures are indicative of recent mass transfer. \cite{Gosnell2019} subsequently measured the masses and effective temperatures of two WD companions in NGC 188  with the Cosmic Origins Spectrograph (COS) on board HST, one of which is in a long-period orbit ($P_{\mathrm{orb}} = 3030$~d; \citealt{Geller2009}) around the BSS WOCS 4540. The authors determined the mass of the long-period white dwarf to be $0.53\pm0.03$~\Msolar, consistent with it being a CO white dwarf with a cooling age of 105 Myr. Ultraviolet photometric observations of M67 and the younger open cluster NGC 7789 (1.6 Gyr, 1.8~\Msolar; \citealt{Gim1998}) have also been conducted with the Ultra-Violet Imaging Telescope on board AstroSat (\citealt{Agrawal2006, Kumar2012}). These observations indicate the presence of ultraviolet excesses in 36$\pm$16\% and 33$\pm$15\% (our Poisson uncertainties) of the BSS populations of these two clusters, respectively (\citealt{Pandey2021, Vaidya2022}). The presence of ultraviolet excesses suggests the presence of hot WD companions, again indicating recent mass transfer among the BSSs of these open clusters.

If mass transfer from an AGB companion were responsible for the creation of some BSSs in open clusters, it may leave behind distinctive chemical signatures in the form of enhancements in \textit{s}-process elements such as barium (Ba), which are produced during the thermally-pulsing phase of AGB evolution (\citealt{Iben1983, Busso1999}). Such enhancements in Ba have been observed among field binaries, both among giants (\citealt{Bidelman1951, McClure1983, Zacs1997, Udry1998, Shetye2018, Jorissen2019}) and dwarfs (\citealt{Tomkin1989, Lu1991, North1991, Gray2011, Escorza2017, Escorza2019}). Evidence of previous extrinsic \textit{s}-process enhancement through mass transfer has also been identified in field binaries containing AGB stars (\citealt{Shetye2020, Shetye2021}).

Comparatively little work, however, has been done to identify Ba enrichment in open clusters. Two Ba-enriched giant stars were identified in the open cluster NGC 2420 (2.4 Gyr, 1.6~\Msolar; \citealt{Demarque1994}) by \cite{Smith1987}, and another two Ba-enriched giants were identified in NGC 5822 (900 Myr, 2.6~\Msolar; \citealt{Carraro2011}) by \cite{KS2013}. \cite{Mathys1991} conducted a spectroscopic survey of the BSSs of M67 and identified two that were significantly enriched in Ba (see Section \ref{subsec:amount}). \cite{McGahee2014} studied the population of yellow stragglers in M67 and found that none were enriched in \textit{s}-process elements.

The first dedicated Ba abundance survey of BSSs in open clusters was conducted by \citet[hereafter \citetalias{Milliman2015}]{Milliman2015}, who studied the BSSs in the open cluster NGC 6819 (2.5 Gyr, 1.5~\Msolar; \citealt{Yang2013}). The authors found five Ba-enriched BSSs compared to the MS, indicative of mass transfer from an AGB companion. Remarkably, four of these five Ba-enriched BSSs were not observed to be radial-velocity-variable and the fifth was a double-lined system (see Section \ref{subsec:binarity}). Subsequent spectroscopic observations of the BSSs of NGC 188 conducted by \cite{Milliman2016} revealed that none of the BSSs were enriched in Ba compared to the MS.

In this work we extend the Ba abundance survey of BSSs in open clusters as part of the ongoing WIYN Open Cluster Study (WOCS; \citealt{Mathieu2000}), focusing on the BSSs of NGC 7789 and M67. We present the results of our study of the BSSs of these clusters in this work, and we combine our results with those of \citetalias{Milliman2015} and \cite{Milliman2016} to study the connection between the initial mass of the AGB donor and the observed properties of the BSSs.

\vspace{-0.67em}

\section{Targets and Observations}
\label{sec:targets}

We selected member stars in M67 based on the radial-velocity (RV) membership analysis of \cite{Geller2015}. The authors identified a total of 14 BSS members, 13 of which we observed in this study. We did not observe the BSS member WOCS 1010 (S977), as previous observations have indicated that this star is too hot and too rapidly rotating for reliable abundance analysis using our methodology (see, e.g. \citealt{Landsman1998, BM2018}). We also observed seven of the 11 blue-lurker member stars as described in \cite{Leiner2019}, two yellow stragglers \citep[out of four;][]{Mathieu1986, Mathieu1990, Geller2015}, and one sub-subgiant \citep[out of two;][]{Mathieu2003}. Table \ref{tab:star_params} presents the photometric information and stellar parameters for each star observed in this study. 

We selected member stars in NGC 7789 in a similar manner, drawing from the RV and proper motion-membership analysis of \cite{Nine2020}. We observed 10 of the 12 identified BSS members in NGC 7789, and the photometric information and stellar parameters for the stars observed in this cluster are listed in Table \ref{blue:tab:ngc7789_star_params}. 

We further selected as control samples MS stars in both clusters that are not velocity variable (i.e., that are likely single stars), and that are not too rapidly rotating for reliable abundance analysis (\vsini~$<10$~\kms~in M67, \vsini~$\lesssim40$~\kms~in NGC 7789). We list the photometric information and stellar parameters for these stars as well in Tables \ref{tab:star_params} and \ref{blue:tab:ngc7789_star_params}.

We observed the M67 stellar sample with the Hydra Multi-Object Spectrograph (MOS; \citealt{Barden1994}) on the WIYN\footnote{The WIYN 3.5m Observatory is a joint facility of the University of Wisconsin–Madison, Indiana University, NSF’s NOIRLab, the Pennsylvania State University and Purdue University.} 3.5m telescope. We conducted observations over two nights in 2022 February. We observed the stellar sample in two spectral orders, one order centered on 5890~\r{A} and spanning 5750--6000~\r{A} (6.7~hr, $R\sim17,000$), and the other centered on 6750~\r{A} and spanning 6550--6950~\r{A} (7~hr, $R\sim17,000$). We conducted observations for the NGC 7789 stellar sample in 2022 October in the same two spectral orders and with similar observation times (7.3~hr in the first order, and 10.3 hr in the second). Spectra gathered in both 2022 February and in 2022 October have typical signal-to-noise ratios $>100$.

All spectra were reduced using IRAF (\citealt{Tody1986}), following the procedure described in \cite{Steinhauer2003} and incorporating the L.A. Cosmic routine for improved cosmic ray rejection (\citealt{vanDokkum2001}). We refer the reader to \citetalias{Milliman2015} for more detail about our reduction procedure. Our results for Ba abundances in this work are derived from the 5853.7~\r{A} Ba II feature. The iron abundances for each star were derived from Fe I features in the spectral orders observed in both spectral orders.

\section{Atmospheric Parameters}
\label{sec:atm_params}

We followed the procedure of \cite{Nine2023} to derive stellar atmospheric parameters, which we summarize here. We present the atmospheric parameters in Tables \ref{tab:star_params} and \ref{blue:tab:ngc7789_star_params}.

We used Python packages to derive the stellar effective temperatures by first computing continuum-normalized \halpha\, spectra using \texttt{specutils} (\citealt{Earl2022}). We then construct a grid of theoretical solar-metallicity \halpha\, profiles\footnote{\url{https://github.com/barklem/public-data/tree/master/hydrogen-line-grids}} obtained from \cite{Barklem2002}, spanning $3.6 \leq $~\logg~$ \leq 4.5$ and $5000 \leq$~\teff$~\leq 7500$~K. We estimated the \logg\, of each star with the \texttt{starfit} function of \texttt{isochrones} (\citealt{Morton2015}), which interpolates across a grid of MIST isochrones (\citealt{Choi2016, Dotter2016}). 

We then derived the \vsini~ of each star following the method of \cite{Rhode2001}. Our minimum velocity resolution is $\sim$10~\kms~(\citealt{Rhode2001, Geller2009}), and we do not report \vsini~ values below this limit in Tables \ref{tab:star_params} and \ref{blue:tab:ngc7789_star_params}. We then rotationally broaden the theoretical \halpha\, profiles using the convolution method of \cite{Gray2005} as implemented in \texttt{PyAstronomy}\footnote{\url{https://pyastronomy.readthedocs.io/en/latest/}} (\citealt{Czesla2019}) to match the measured \vsini\, of each star. For those stars with \vsini\, less than 10 \kms, we rotationally broadened their spectra to 10 \kms. We then found the best-fit \teff\, and uncertainties through $\chi^2$-minimization of the wings of the broadened \halpha\, profiles with respect to the observed spectra. The derived values of \teff\, and their uncertainties are listed in Tables \ref{tab:star_params} and \ref{blue:tab:ngc7789_star_params}. For the stars whose effective temperatures or \logg\, fall outside of the \cite{Barklem2002} grid, we followed the same procedure, instead using theoretical solar-metallicity spectra\footnote{\url{http://specmodels.iag.usp.br/fits_search/}} from \cite{Coelho2014}. We extracted the spectral region around \halpha\, and normalized them to compare with our observed spectra. 

The precise values of microturbulence in these atmospheres are unknown. We varied the microturbulence velocity (\vt) parameter based on the results of \cite{Takeda2008}, \cite{Ramirez2013}, and \citetalias{Milliman2015} as follows: for the coolest stars in our sample (\teff~$\leq 6200$~K), we varied \vt\, between 1.0--1.5 \kms; for stars with $6200 <$~\teff~$<7400$~K, we varied \vt\, between 2.0--2.5 \kms; for stars with $7400 \leq$~\teff~$<8000$~K, we varied \vt\, between 2.0--3.0 \kms; and for stars with \teff~$\geq 8000$~K, we varied \vt\, between 1.0--4.0 \kms. The ranges of \vt\, which we adopted for each star are listed in Tables \ref{tab:star_params} and \ref{blue:tab:ngc7789_star_params}.   

\centerwidetable
\begin{deluxetable*}{ccccccccccc}[p!]
\tabletypesize{\footnotesize}
\tablecaption{M67 Stellar Parameters\label{tab:star_params}}
\tablehead{\colhead{WOCS ID} & \colhead{Sanders ID\tablenotemark{a}} & $\alpha$ (J2000) & $\delta$ (J2000) & \colhead{$G$\tablenotemark{b}} & \colhead{$G_{\mathrm{BP}}-G_{\mathrm{RP}}$\tablenotemark{b}}  & \colhead{$T_{\mathrm{eff}}$} & \colhead{log$_{10}(g)$} & \colhead{$v_t$}  & \colhead{$v$sin$i$} & \colhead{$P_{\mathrm{orb}}$} \\
\colhead{} & \colhead{} &  \colhead{} & \colhead{} & \colhead{} & \colhead{} & \colhead{(K)} & \colhead{} & \colhead{(km s$^{-1}$)}  & \colhead{(km s$^{-1}$)} & \colhead{(d)}}
\tablewidth{0pt}
\startdata
\multicolumn{10}{l}{\textit{Blue Stragglers\tablenotemark{c}:}}\\
1006\tablenotemark{*} & 1066 & 8 51 27.01 & 11 51 52.6 & 10.93 & 0.12 & 8210$^{+140}_{-110}$ & 4.05$^{+0.04}_{-0.03}$ & 1.0$-$4.0 & 85 & \nodata \\
1007\tablenotemark{*} & 1284 & 8 51 34.31 & 11 51 10.5 & 10.89 & 0.34 & 7420$^{+80}_{-80}$ & 3.87$^{+0.03}_{-0.03}$ & 2.0$-$3.0 & 85 & 4.18287 \\
1025\tablenotemark{*} & 1195 & 8 51 37.69 & 11 37 03.8 & 12.22 & 0.54 & 6590$^{+70}_{-60}$ & 4.07$^{+0.03}_{-0.03}$ & 2.0$-$2.5 & 65 & 1154 \\
1026\tablenotemark{*} & 1434 & 8 52 10.74 & 11 44 05.9 & 10.64 & 0.14 & 8220$^{+150}_{-100}$ & 3.95$^{+0.03}_{-0.03}$ & 1.0$-$4.0 & $>$140 & \nodata \\
2009 & 1082 & 8 51 20.79 & 11 53 26.1 & 11.62 & 0.58 & 6550$^{+60}_{-50}$ & 3.69$^{+0.03}_{-0.02}$ & 2.0$-$2.5 & 17$^{+2}_{-3}$ & 1.0677978, 1189 \\
2011 &  968 & 8 51 26.43 & 11 43 50.7 & 11.25 & 0.15 & 8280$^{+320}_{-190}$ & 4.12$^{+0.04}_{-0.03}$ & 1.0$-$4.0 & $<$10 & \nodata \\
2013 & 1267 & 8 51 48.64 & 11 49 15.6 & 10.87 & 0.28 & 7840$^{+150}_{-140}$ & 3.73$^{+0.03}_{-0.03}$ & 2.0$-$3.0 & 67$^{+2}_{-2}$ & 846 \\
3005 & 1263 & 8 51 32.59 & 11 48 52.1 & 11.03 & 0.28 & 7790$^{+150}_{-140}$ & 3.95$^{+0.03}_{-0.03}$ & 2.0$-$3.0 & 23.6$^{+0.5}_{-0.5}$ & \nodata \\
3010 &  975 & 8 51 14.36 & 11 45 00.5 & 10.98 & 0.58 & 6440$^{+70}_{-70}$ & 3.64$^{+0.02}_{-0.06}$ & 2.0$-$2.5 & 47$^{+2}_{-2}$ & 1221 \\
3013\tablenotemark{*} &  752 & 8 51 03.52 & 11 45 02.8 & 11.26 & 0.39 & 7340$^{+100}_{-90}$ & 3.90$^{+0.03}_{-0.03}$ & 2.0$-$3.0 & 70 & 1003 \\
4006\tablenotemark{*} & 1280 & 8 51 32.58 & 11 50 40.6 & 12.18 & 0.34 & 7560$^{+80}_{-80}$ & 4.27$^{+0.03}_{-0.03}$ & 2.0$-$3.0 & 120 & \nodata \\
5005 &  997 & 8 51 19.90 & 11 47 00.4 & 12.02 & 0.62 & 6490$^{+70}_{-70}$ & 3.90$^{+0.03}_{-0.03}$ & 2.0$-$2.5 & 16.9$^{+0.8}_{-0.8}$ & 4913 \\
6038 & 2226 & 8 51 28.36 & 12 07 38.5 & 12.45 & 0.60 & 6470$^{+70}_{-80}$ & 4.05$^{+0.03}_{-0.03}$ & 2.0$-$2.5 & 14.7$^{+0.8}_{-0.9}$ & \nodata \\
\\
\multicolumn{10}{l}{\textit{Blue Lurkers\tablenotemark{d}:}}\\
1020 &   751 & 8 50 47.66 & 11 44 51.4 & 12.58 & 0.66 & 6280$^{+100}_{-90}$ & 4.06$^{+0.03}_{-0.03}$ & 2.0$-$2.5 & 15$^{+4}_{-5}$ & \nodata \\
3001 &  1031 & 8 51 22.96 & 11 49 13.1 & 13.15 & 0.63 & 6430$^{+70}_{-80}$ & 4.29$^{+0.03}_{-0.03}$ & 2.0$-$2.5 & 24.7$^{+0.6}_{-0.6}$ & 128.14 \\
6025 &  1431 & 8 52 05.81 & 11 42 24.7 & 13.56 & 0.78 & 6430$^{+70}_{-80}$ & 4.23$^{+0.03}_{-0.03}$ & 2.0$-$2.5 & $<$10 & 6265 \\
7035 &   648 & 8 50 42.62 & 12 03 10.2 & 13.22 & 0.72 & 6050$^{+110}_{-90}$ & 4.15$^{+0.03}_{-0.03}$ & 1.0$-$1.5 & $<$10 & \nodata \\ 
9005 &  1005 & 8 51 15.45 & 11 47 31.4 & 12.55 & 0.66 & 6240$^{+100}_{-90}$ & 3.98$^{+0.03}_{-0.03}$ & 2.0$-$2.5 & $<$10 & 2769 \\
12020 & 1102 & 8 51 08.85 & 11 57 53.7 & 14.09 & 0.78 & 5960$^{+100}_{-110}$ & 4.44$^{+0.02}_{-0.03}$ & 1.0$-$1.5 & $<$10 & 762 \\
14020 & 1452 & 8 52 03.50 & 11 47 48.1 & 14.42 & 0.83 & 5900$^{+130}_{-150}$ & 4.45$^{+0.03}_{-0.04}$ & 1.0$-$1.5 & $<$10 & 358.9 \\
\\
\multicolumn{10}{l}{\textit{Yellow Stragglers and Sub-subgiants\tablenotemark{e}:}}\\
1015 &  1237 & 8 51 50.20 & 11 46 07.0 & 10.51 & 1.10 & 5000$^{+80}_{-80}$ & 2.90$^{+0.01}_{-0.02}$ & 1.0$-$1.5 & 12.9$^{+0.9}_{-1.0}$ & 698.56 \\
2008 &  1072 & 8 51 21.76 & 11 52 37.8 & 11.13 & 0.81 & 5920$^{+120}_{-120}$ & 3.45$^{+0.03}_{-0.02}$ & 1.0$-$1.5 & 17.6$^{+0.8}_{-0.8}$ & 1513 \\
13008 & 1063 & 8 51 13.36 & 11 51 40.1 & 13.13 & 1.24 & 4880$^{+150}_{-150}$ & 3.69$^{+0.03}_{-0.03}$ & 1.0$-$1.5 & 13.4$^{+1.2}_{-1.4}$ & 18.38775 \\
\\
\multicolumn{10}{l}{\textit{Main Sequence:}}\\
3024 &   733 & 8 50 53.08 & 11 40 02.2 & 13.52 & 0.74 & 6130$^{+110}_{-110}$ & 4.29$^{+0.04}_{-0.03}$ & 2.0$-$2.5 & $<$10 & \nodata \\
5017 &   803 & 8 50 56.01 & 11 53 52.0 & 13.50 & 0.74 & 6040$^{+90}_{-100}$ & 4.27$^{+0.03}_{-0.03}$ & 1.0$-$1.5 & $<$10 & \nodata \\
5022 &  1464 & 8 52 07.42 & 11 50 22.1 & 13.08 & 0.77 & 5980$^{+110}_{-100}$ & 4.08$^{+0.02}_{-0.04}$ & 1.0$-$1.5 & $<$10 & \nodata \\
7021 &   944 & 8 51 26.48 & 11 38 36.8 & 13.54 & 0.74 & 6070$^{+100}_{-110}$ & 4.31$^{+0.04}_{-0.03}$ & 1.0$-$1.5 & $<$10 & \nodata \\
7031 &   722 & 8 50 57.86 & 11 35 14.8 & 13.93 & 0.78 & 6020$^{+100}_{-130}$ & 4.36$^{+0.04}_{-0.04}$ & 1.0$-$1.5 & $<$10 & \nodata \\
7045 &  1130 & 8 51 12.54 & 12 11 17.4 & 14.22 & 0.83 & 5820$^{+170}_{-130}$ & 4.34$^{+0.04}_{-0.04}$ & 1.0$-$1.5 & $<$10 & \nodata \\
8012 &   967 & 8 51 18.34 & 11 43 25.2 & 13.28 & 0.75 & 6110$^{+90}_{-100}$ & 4.14$^{+0.03}_{-0.03}$ & 1.0$-$1.5 & $<$10 & \nodata \\
8030 &  1491 & 8 51 58.07 & 12 00 59.6 & 13.64 & 0.76 & 6090$^{+100}_{-100}$ & 4.27$^{+0.03}_{-0.03}$ & 1.0$-$1.5 & $<$10 & \nodata \\
10012 &  796 & 8 51 04.93 & 11 52 26.2 & 13.70 & 0.76 & 6050$^{+100}_{-110}$ & 4.39$^{+0.02}_{-0.03}$ & 1.0$-$1.5 & $<$10 & \nodata \\
10014 & 1093 & 8 51 18.72 & 11 55 49.7 & 13.98 & 0.79 & 6060$^{+100}_{-110}$ & 4.32$^{+0.04}_{-0.03}$ & 1.0$-$1.5 & $<$10 & \nodata \\
11015 &  788 & 8 50 56.33 & 11 51 29.3 & 13.96 & 0.79 & 6010$^{+110}_{-110}$ & 4.34$^{+0.04}_{-0.04}$ & 1.0$-$1.5 & $<$10 & \nodata \\
11019 & 1477 & 8 51 53.30 & 11 54 19.5 & 14.41 & 0.87 & 5900$^{+130}_{-140}$ & 4.37$^{+0.04}_{-0.04}$ & 1.0$-$1.5 & $<$10 & \nodata \\
12025 & 1421 & 8 51 53.89 & 11 39 04.8 & 14.35 & 0.83 & 5840$^{+140}_{-130}$ & 4.44$^{+0.03}_{-0.04}$ & 1.0$-$1.5 & $<$10 & \nodata \\
12028 &  597 & 8 50 42.50 & 11 39 49.3 & 13.28 & 0.74 & 6100$^{+110}_{-110}$ & 4.19$^{+0.03}_{-0.03}$ & 2.0$-$2.5 & $<$10 & \nodata \\
16011 &  770 & 8 51 00.81 & 11 48 52.8 & 14.45 & 0.86 & 5740$^{+150}_{-160}$ & 4.46$^{+0.03}_{-0.04}$ & 1.0$-$1.5 & $<$10 & \nodata \\
\enddata
\tablenotetext{*}{These are the stars for which we derived \vsini\, and abundances solely through spectral synthesis (see Section \ref{sec:abun_analysis}).}
\tablenotetext{a}{\cite{Sanders1977}}
\tablenotetext{b}{\textit{Gaia} DR3; \cite{GAIA2016, GaiaDR3}}
\tablenotetext{c}{\cite{Milone1991, Latham1996, Sandquist2003, Geller2015}}
\tablenotetext{d}{\cite{Leiner2019, Geller2021, Nine2023}}
\tablenotetext{e}{\cite{Mathieu1986, Mathieu1990, Mathieu2003, Geller2015, Geller2021}}
\end{deluxetable*}%
\begin{deluxetable*}{cccccccccc}
\tabletypesize{\footnotesize}
\tablecaption{NGC 7789 Stellar Parameters\label{blue:tab:ngc7789_star_params}}
\tablehead{\colhead{WOCS ID}  & \colhead{$\alpha$ (J2000)} & \colhead{$\delta$ (J2000)} & \colhead{$G$\tablenotemark{a}} & \colhead{$G_{\mathrm{BP}}-G_{\mathrm{RP}}$\tablenotemark{a}}  & \colhead{$T_{\mathrm{eff}}$} & \colhead{log$_{10}(g)$} & \colhead{$v_t$} & \colhead{$v$sin$i$} & \colhead{$P_{\mathrm{orb}}$\tablenotemark{b}} \\
\colhead{} & \colhead{} & \colhead{} & \colhead{} & \colhead{} & \colhead{(K)} & \colhead{} & \colhead{(km s$^{-1}$)}  & \colhead{(km s$^{-1}$)} & \colhead{(d)}}
\tablewidth{0pt}
\startdata
\multicolumn{10}{l}{\textit{Blue Stragglers:}}\\
5004  & 23 57 30.39 & 56 44 30.3 & 12.76 & 0.55 & 7870$^{+330}_{-190}$ & 3.64$^{+0.06}_{-0.05}$ & 1.0--4.0 & 16.4$^{+1.3}_{-1.4}$ & \nodata\\
5011\tablenotemark{*}  & 23 56 59.03 & 56 47 33.2 & 12.85 & 0.54 & 8350$^{+140}_{-130}$ & 3.66$^{+0.07}_{-0.06}$ & 1.0--4.0 & 50 & 2710\\
10010\tablenotemark{*}& 23 56 54.24 & 56 40 46.9 & 13.71 & 0.53 & 8360$^{+110}_{-100}$ & 3.92$^{+0.09}_{-0.09}$ & 1.0--4.0 & 35 & \nodata\\
10011 & 23 56 42.13 & 56 44 12.4 & 13.48 & 0.75 & 7150$^{+270}_{-170}$ & 3.67$^{+0.07}_{-0.06}$ & 2.0--3.0 & 29.8$^{+1.9}_{-1.9}$ & 517\\
15015\tablenotemark{*} & 23 58 05.25 & 56 47 02.8 & 13.86 & 0.53 & 8300$^{+160}_{-170}$ & 3.96$^{+0.09}_{-0.09}$ & 1.0--4.0 & 50 & \nodata\\
16020 & 23 58 17.80 & 56 49 40.0 & 13.79 & 0.63 & 7820$^{+140}_{-120}$ & 3.84$^{+0.09}_{-0.09}$ & 2.0--3.0 & 21.7$^{+2.3}_{-2.5}$ & \nodata\\
20009\tablenotemark{*} & 23 57 49.78 & 56 44 37.0 & 14.31 & 0.62 & 7910$^{+180}_{-100}$ & 4.00$^{+0.10}_{-0.09}$ & 1.0--4.0 & 30 & 4190\\
25008\tablenotemark{*} & 23 57 30.94 & 56 39 49.1 & 13.96 & 0.78 & 7280$^{+140}_{-110}$ & 3.77$^{+0.08}_{-0.08}$ & 2.0--3.0 & 45 & \nodata\\
25024\tablenotemark{*} & 23 58 45.70 & 56 44 39.9 & 14.78 & 0.76 & 7330$^{+130}_{-120}$ & 4.02$^{+0.10}_{-0.10}$ & 2.0--3.0 & 60 & \nodata\\
36011 & 23 57 08.83 & 56 38 37.5 & 14.65 & 0.73 & 7550$^{+100}_{-120}$ & 4.01$^{+0.09}_{-0.10}$ & 2.0--3.0 & 20.2$^{+1.0}_{-1.0}$ & 217.6\\
\\
\multicolumn{10}{l}{\textit{Main Sequence:}}\\
5001  & 23 57 19.84 & 56 43 36.2 & 14.70 & 0.85 & 6650$^{+140}_{-160}$ & 3.92$^{+0.09}_{-0.09}$ & 2.0--2.5 & 17.9$^{+1.1}_{-1.1}$ & \nodata\\
13015 & 23 57 48.28 & 56 36 50.4 & 13.43 & 0.92 & 6540$^{+110}_{-110}$ & 3.48$^{+0.09}_{-0.09}$ & 2.0--2.5 & 33.4$^{+2.0}_{-2.0}$ & \nodata\\
14005 & 23 57 33.29 & 56 41 35.9 & 13.92 & 0.90 & 6700$^{+40}_{-40}$ & 3.67$^{+0.08}_{-0.09}$ & 2.0--2.5 & 23.9$^{+2.2}_{-2.4}$ & \nodata\\
17016 & 23 57 02.70 & 56 35 53.3 & 13.91 & 0.89 & 6430$^{+150}_{-160}$ & 3.67$^{+0.08}_{-0.08}$ & 2.0--2.5 & 45$^{+3}_{-3}$ & \nodata\\
18006 & 23 57 19.13 & 56 40 40.4 & 13.97 & 0.86 & 6680$^{+120}_{-110}$ & 3.71$^{+0.08}_{-0.08}$ & 2.0--2.5 & 24.6$^{+1.5}_{-1.6}$ & \nodata\\
18007 & 23 57 22.88 & 56 46 40.0 & 14.17 & 0.94 & 6560$^{+140}_{-150}$ & 3.70$^{+0.08}_{-0.08}$ & 2.0--2.5 & 36.4$^{+0.4}_{-0.4}$ & \nodata\\
23016 & 23 56 30.90 & 56 47 19.2 & 14.10 & 0.96 & 6410$^{+130}_{-150}$ & 3.67$^{+0.08}_{-0.09}$ & 2.0--2.5 & 29.3$^{+1.5}_{-1.6}$ & \nodata\\
24008 & 23 57 36.62 & 56 40 08.6 & 13.88 & 0.95 & 6270$^{+150}_{-160}$ & 3.61$^{+0.10}_{-0.11}$ & 2.0--2.5 & 45$^{+3}_{-3}$ & \nodata\\
25016 & 23 56 52.41 & 56 50 15.8 & 14.28 & 0.92 & 6590$^{+140}_{-160}$ & 3.74$^{+0.08}_{-0.08}$ & 2.0--2.5 & 17.7$^{+0.7}_{-0.8}$ & \nodata\\
26009 & 23 57 46.64 & 56 41 06.0 & 14.44 & 0.94 & 6550$^{+130}_{-140}$ & 3.77$^{+0.08}_{-0.08}$ & 2.0--2.5 & 36.6$^{+1.2}_{-1.2}$ & \nodata\\
27007 & 23 57 42.97 & 56 44 50.0 & 14.52 & 0.93 & 6650$^{+150}_{-160}$ & 3.81$^{+0.08}_{-0.08}$ & 2.0--2.5 & 26.9$^{+1.5}_{-1.5}$ & \nodata\\
28015 & 23 56 42.87 & 56 38 28.8 & 14.53 & 0.87 & 6650$^{+160}_{-80}$  & 3.85$^{+0.09}_{-0.09}$ & 2.0--2.5 & 17$^{+4}_{-5}$ & \nodata\\
33011 & 23 57 06.29 & 56 47 54.7 & 14.49 & 0.90 & 6640$^{+150}_{-150}$ & 3.82$^{+0.08}_{-0.09}$ & 2.0--2.5 & 15$^{+3}_{-3}$ & \nodata\\
37011 & 23 57 28.61 & 56 48 22.9 & 14.64 & 0.87 & 6600$^{+110}_{-120}$ & 3.89$^{+0.08}_{-0.08}$ & 2.0--2.5 & 37.3$^{+1.0}_{-1.0}$ & \nodata\\
37016 & 23 57 15.92 & 56 50 57.8 & 14.76 & 0.91 & 6650$^{+150}_{-150}$ & 3.88$^{+0.09}_{-0.09}$ & 2.0--2.5 & 22.2$^{+0.7}_{-0.7}$ & \nodata\\
42011 & 23 57 55.85 & 56 41 38.7 & 14.79 & 0.98 & 6340$^{+170}_{-190}$ & 3.83$^{+0.09}_{-0.08}$ & 2.0--2.5 & 16.1$^{+0.4}_{-0.4}$ & \nodata\\
42016 & 23 56 27.89 & 56 45 53.2 & 14.78 & 0.95 & 6540$^{+150}_{-140}$ & 3.86$^{+0.08}_{-0.08}$ & 2.0--2.5 & 21.8$^{+0.6}_{-0.6}$ & \nodata\\
\enddata
\tablenotetext{*}{These are the stars for which we derived \vsini\, and abundances solely through spectral synthesis (see Section \ref{sec:abun_analysis}).}
\tablenotetext{a}{Gaia DR3; \cite{GAIA2016, GaiaDR3}}
\tablenotetext{b}{\cite{Nine2020}}
\end{deluxetable*}%
\vspace{-5.25em}
\section{Abundance Analysis}
\label{sec:abun_analysis}

Our abundance analyses were performed with the 2019 November version of MOOG\footnote{\url{https://www.as.utexas.edu/~chris/moog.html}} (\citealt{Sneden1973}), as implemented in \texttt{pyMOOGi}\footnote{\url{https://github.com/madamow/pymoogi}} (\citealt{Adamow2017}). We computed model atmospheres for each combination of \teff\, and \vt\, for every star using Version 9 of the \texttt{ATLAS} code (\citealt{Kurucz1970, Sbordone2004, Castelli2005ATLAS, Kurucz2014}), as implemented in \texttt{BasicATLAS}\footnote{\url{https://github.com/Roman-UCSD/BasicATLAS}} (\citealt{Gerasimov2022, Larkin2023}). We created line lists for each spectral region using atomic information from the third data release of the Vienna Atomic Line Database\footnote{\url{http://vald.astro.uu.se/}} (VALD3; \citealt{Ryabchikova2015}).

We measured equivalent widths (EWs) for strong, isolated Fe I lines in our spectral regions with SPECTRE\footnote{\url{https://www.as.utexas.edu/~chris/spectre.html}}. We then use the MOOG task \texttt{abfind} with our measured EWs to compute A(Fe)\footnote{A(X) = 12.0 + log$_{10}\left(N_\mathrm{X}/N_\mathrm{H}\right)$} for each star. We then calculate [Fe/H]\footnote{$[\mathrm{X}/\mathrm{H}] = \log_{10}\left(N_\mathrm{X}/N_\mathrm{H}\right)_\star - \log_{10}\left(N_\mathrm{X}/N_\mathrm{H}\right)_\odot$.} by taking the difference between the derived A(Fe)$_\star$ and our adopted A(Fe)$_\odot = 7.50$ (\citealt{Asplund2009}). The number of Fe I lines ($N$) used to determine the [Fe/H] of each star is listed in Tables \ref{tab:ba_abuns} and \ref{tab:ngc7789_ba_abuns}.

In order to compute A(Ba) for the MS sample and the slowly rotating BSSs, we use Version 9 of the curve-of-growth program \texttt{WIDTH}\footnote{\url{https://wwwuser.oats.inaf.it/castelli/sources/width9.html}} (\citealt{Castelli2005WIDTH, Kurucz2005}). We measure the EW of Ba II 5853.7 \r{A} for each star with SPECTRE. We then compute a curve of growth for the same Ba II line for every combination of \teff\, and \logg, and we derive A(Ba) for each star by interpolating each curve of growth at the measured EW. We then compute the Ba abundance for each star relative to iron through the relation:

\begin{equation}
    [\mathrm{Ba}/\mathrm{Fe}] = \left(\mathrm{A(Ba)}_\star - \mathrm{A(Ba)}_\odot\right) - [\mathrm{Fe}/\mathrm{H}],
\end{equation}

\noindent where we adopt A(Ba)$_\odot = 2.18$ (\citealt{Asplund2009}).

In the case of very hot (\teff~$\gtrsim$~8000~K) or rapidly rotating BSSs (\vsini~$\gtrsim$~45~\kms), for which EW measurements are unreliable, we instead derive [Fe/H] and [Ba/Fe] through spectral synthesis with the MOOG task \texttt{synth}. For these stars we find the best-fit \vsini\, to their observed broadening, and we adjust the abundances of Ba and Fe until we find the best match to the observed spectrum. We list the best-fit \vsini\, values in Tables \ref{tab:star_params} and \ref{blue:tab:ngc7789_star_params}, and we include their best-fit abundance measurements in Tables \ref{tab:ba_abuns} and \ref{tab:ngc7789_ba_abuns}. In order to verify the robustness of this alternative method, we applied this same procedure to a representative sample of those stars which are not rapidly rotating or are cooler than 8000~K. We found that the abundances obtained in the two methods are consistent to within $\sim$0.05 dex.

\section{Abundance Results}
\label{sec:results}

\begin{figure*}[htb]
    \gridline{\fig{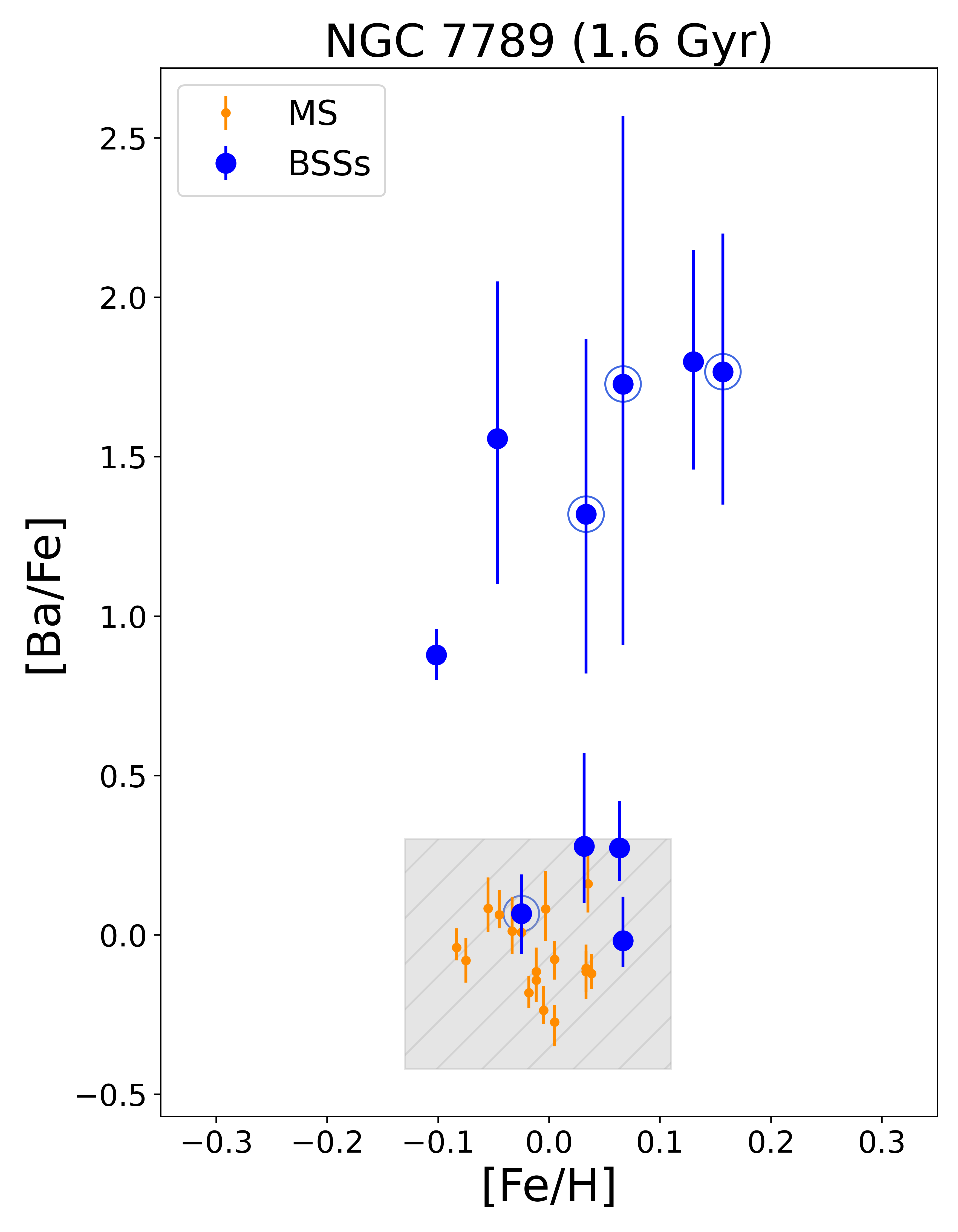}{0.45\linewidth}{}
    \fig{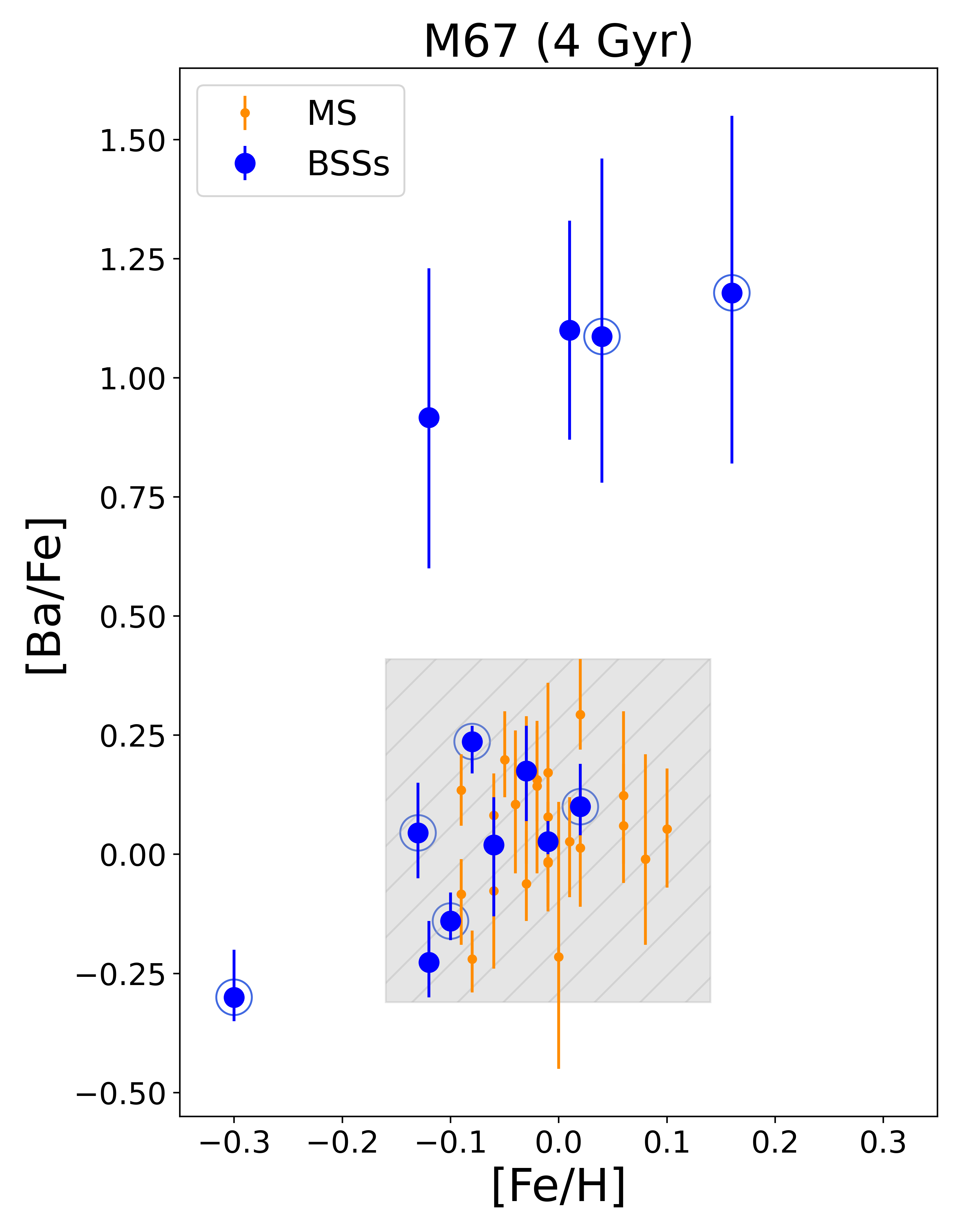}{0.45\linewidth}{}}
    \caption{[Ba/Fe] vs. [Fe/H] diagrams of NGC 7789 (left) and M67 (right). BSSs are indicated with blue points, and the MS control samples are marked with orange points. Binary BSSs are indicated with an additional blue circle. The error bars represent the range of possible [Ba/Fe] values given assumptions about stellar \teff\, and \vt, and the gray hatched boxes represent the 3$\sigma$ uncertainties on the MS [Fe/H] and [Ba/Fe].}
    \label{fig:bafe_plot}
\end{figure*}

In a similar manner as \citetalias{Milliman2015}, for each star we create model atmospheres corresponding to three values of \teff, the best-fit value as well as the $\pm $1$\sigma$ values, and the extrema of assumed \vt\, for each star (see Tables \ref{tab:star_params} and \ref{blue:tab:ngc7789_star_params}). We thus derive six values for each of [Fe/H] and [Ba/Fe]. In Tables \ref{tab:ba_abuns} and \ref{tab:ngc7789_ba_abuns} we present the mean of each of the six measurements for every star, along with the range of measurements determined by the bounds of \teff\, and \vt. 

Based on our sample of MS stars, we find in M67 a MS [Fe/H]~$= -0.01 \pm 0.05$ (s.d.). This value is consistent with previous estimates of the MS metallicity of M67, which range from $-0.05$ to $+0.05$ (see \citealt{Boesgaard2020} and references therein). For NGC 7789 we find a MS [Fe/H]~$= -0.01 \pm 0.04$ (s.d.), consistent with previous measurements of the MS [Fe/H] of NGC 7789 which range from $-0.04$ to $+0.03$ (see \citealt{Overbeek2015} and references therein). 

We list in Table \ref{tab:avg_abun} our measured mean abundances for the MS stars in our sample compared to results from the literature. The literature [Ba/Fe] values from \cite{Pancino2010} and \cite{Overbeek2015} are significantly higher than our measured values, as well as the [Ba/Fe] values for M67 as determined by \cite{Tautv2000}, \cite{Yong2005}, and \cite{Liu2019} (see 
Table \ref{tab:avg_abun}). In both cases, the authors find a solar Ba abundance significantly above their adopted baselines; \cite{Pancino2010} determine a value of A(Ba)$_{\odot} = 2.47$ and \cite{Overbeek2015} determine A(Ba)$_{\odot} = 2.48$, compared to their adopted baseline value of A(Ba)$_{\odot} = 2.13$ from \cite{Anders1989}. If the reported cluster [Ba/Fe] values from these previous works are scaled according to their determined A(Ba)$_{\odot}$ as opposed to their adopted baseline A(Ba)$_{\odot}$, their results for cluster [Ba/Fe] become consistent with those determined in this work and the other literature sources as listed in Table \ref{tab:avg_abun}.

In Figure \ref{fig:bafe_plot} we plot for each cluster the abundance values of the MS and BSS samples as orange and blue dots, respectively, with the error bars indicating the range of measurements in [Ba/Fe]. We also plot a gray box showing the 3$\sigma$ range in MS [Fe/H] and [Ba/Fe] abundances.

\begin{figure*}[ht]
  \gridline{\fig{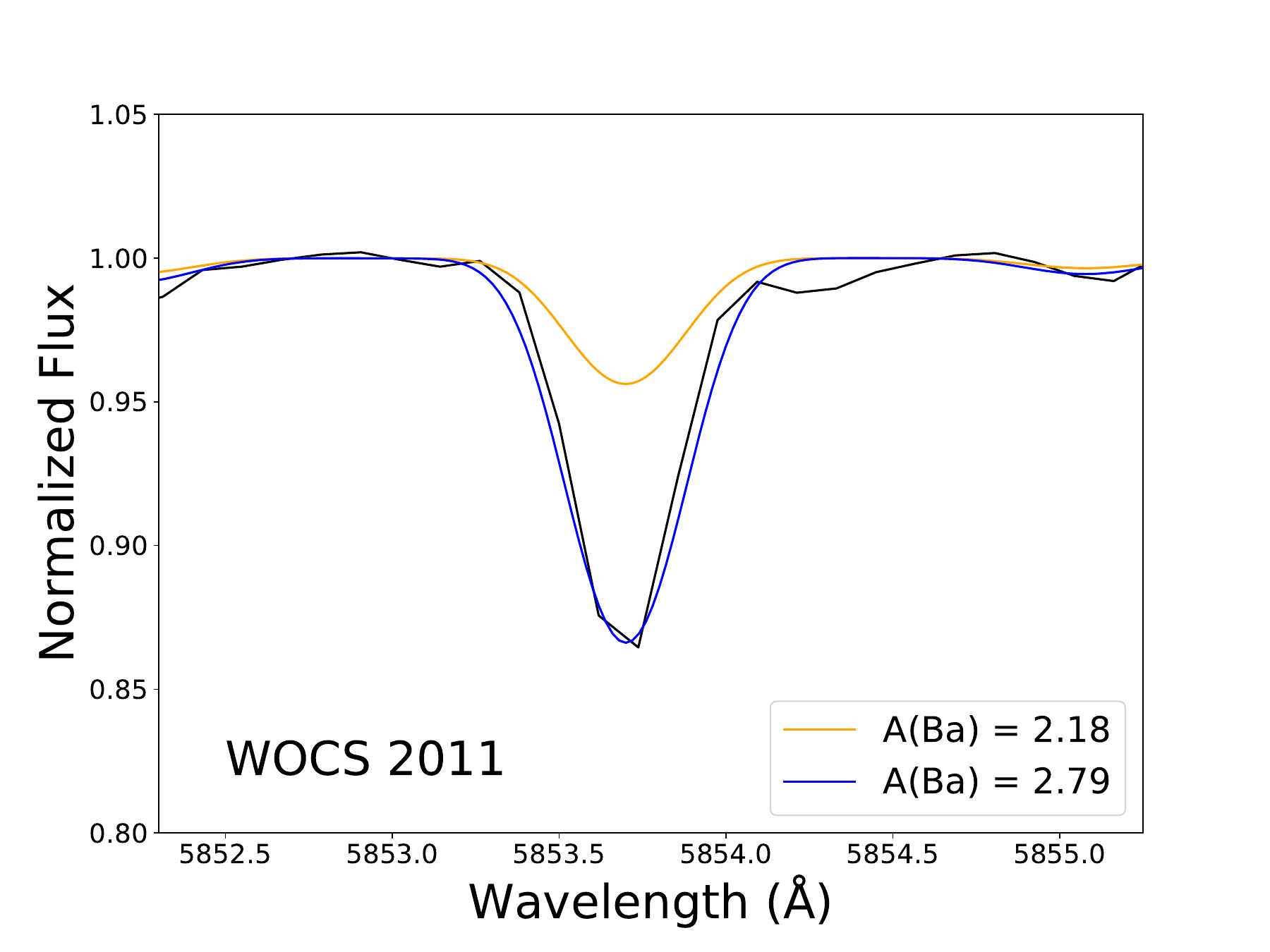}{0.4\linewidth}{}
  \fig{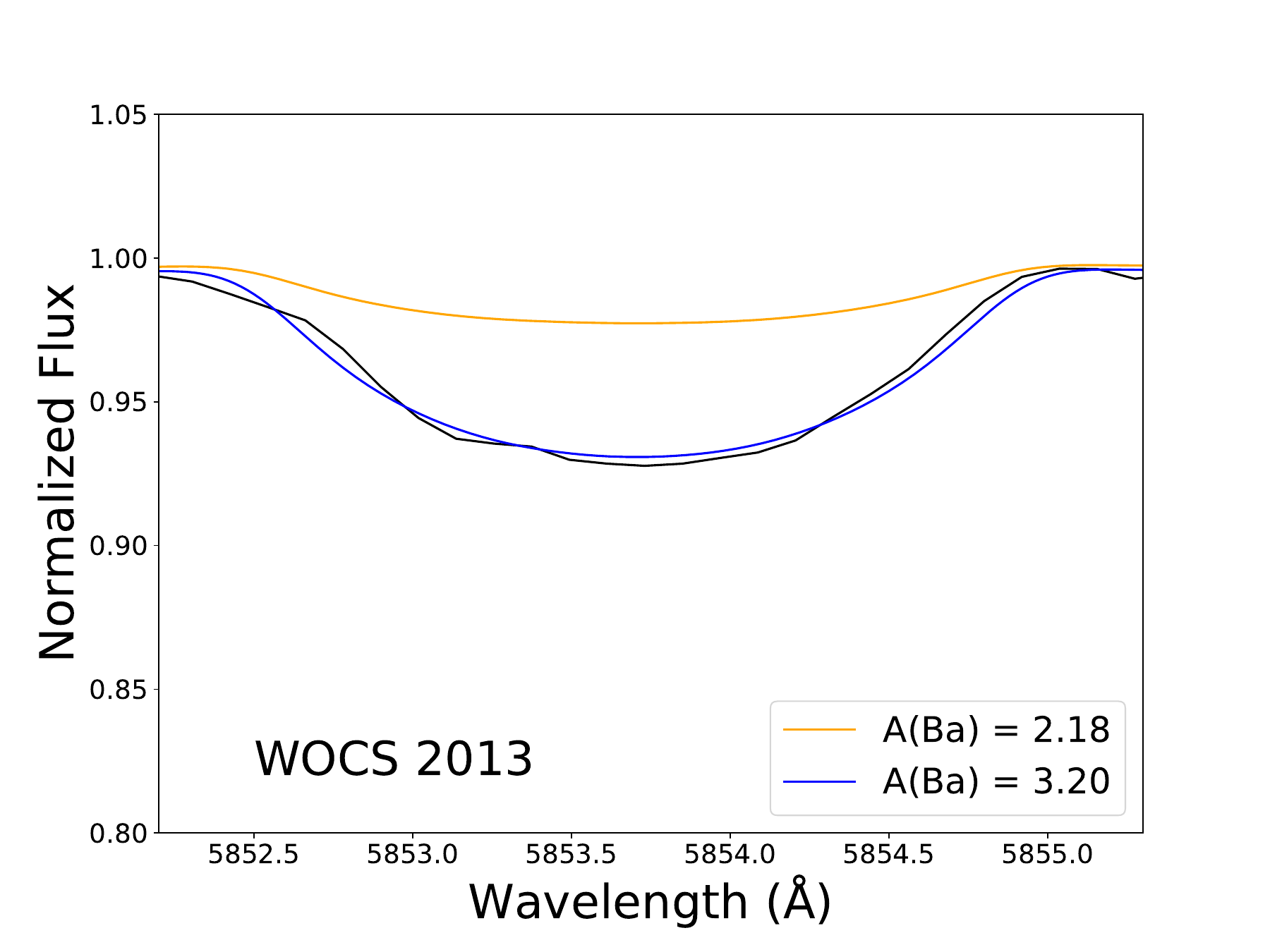}{0.4\linewidth}{}}
  \gridline{\fig{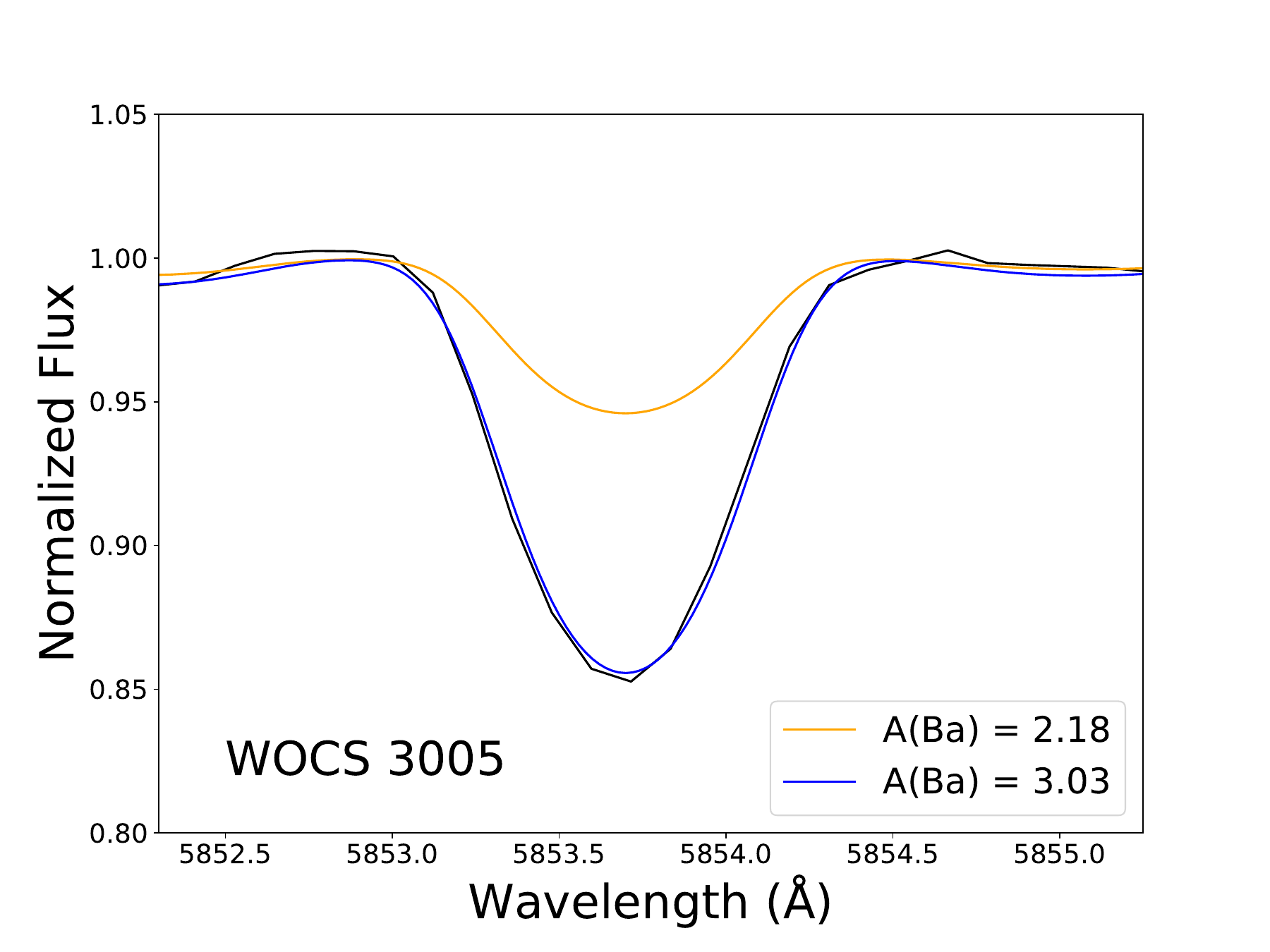}{0.4\linewidth}{}
  \fig{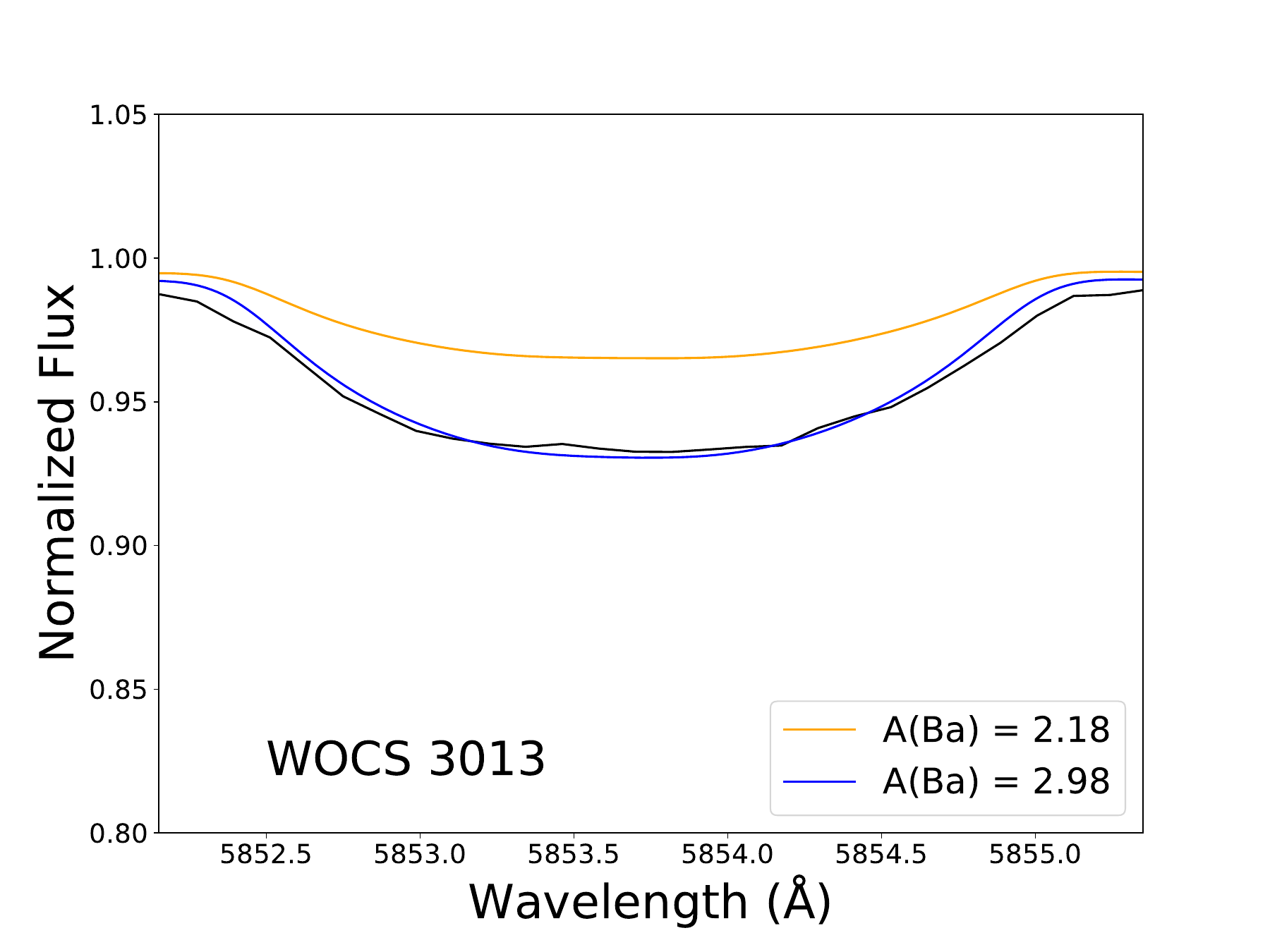}{0.4\linewidth}{}}
  \caption{Observed spectra of the Ba-enriched BSSs in M67 compared to representative model spectra computed with MOOG. We assume the mean \teff\, and maximum \vt\, for each BSS (see Table \ref{tab:star_params}) for these models. In each plot the observed spectrum is plotted in black, the model spectrum with solar A(Ba) is plotted in orange, and the model with the best-fit A(Ba) given the assumed atmospheric parameters is plotted in blue.}
  \label{fig:m67_ba_specs}
\end{figure*}

\begin{figure*}[ht]
  \gridline{\fig{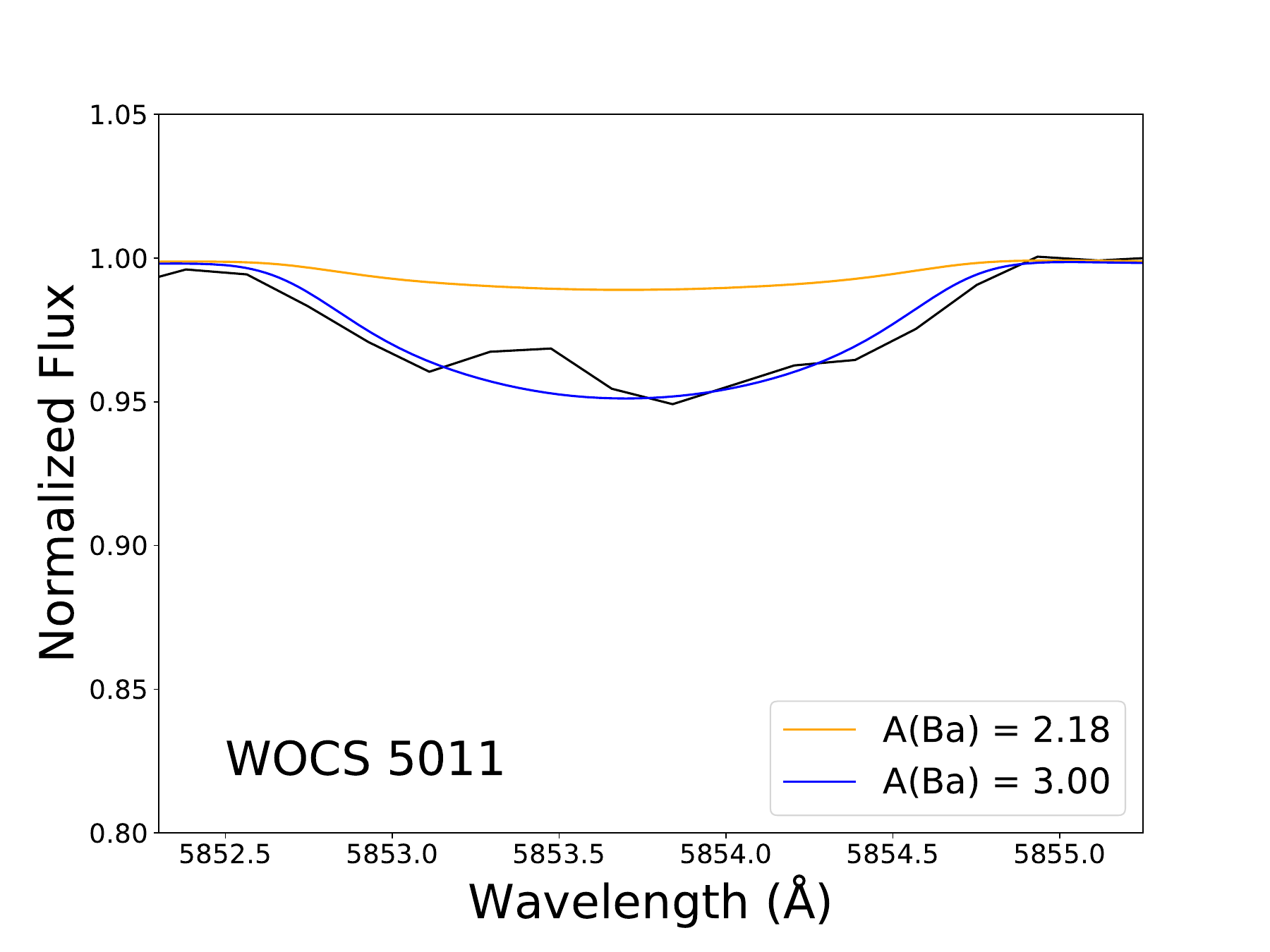}{0.4\linewidth}{}
  \fig{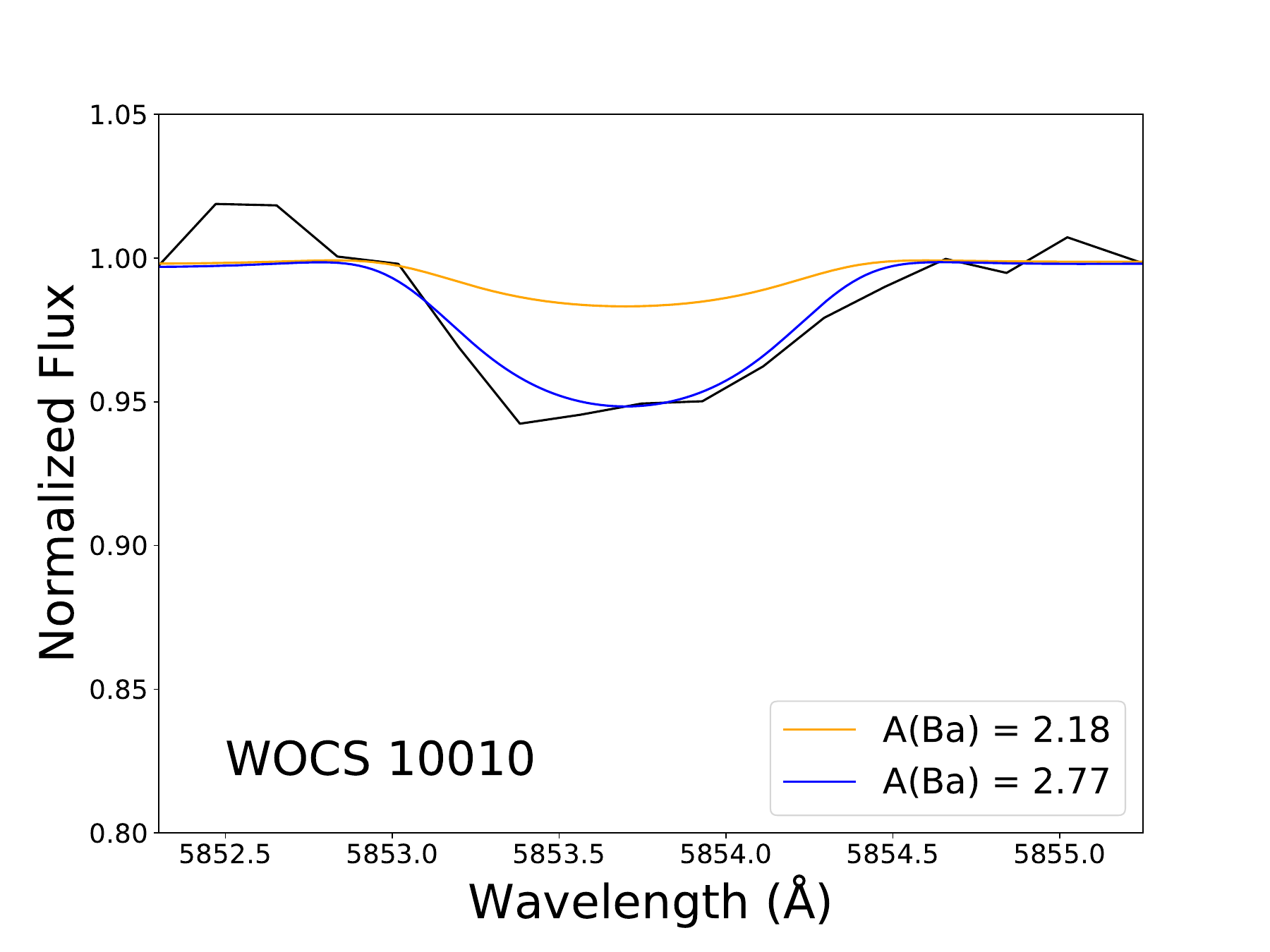}{0.4\linewidth}{}}
  \gridline{\fig{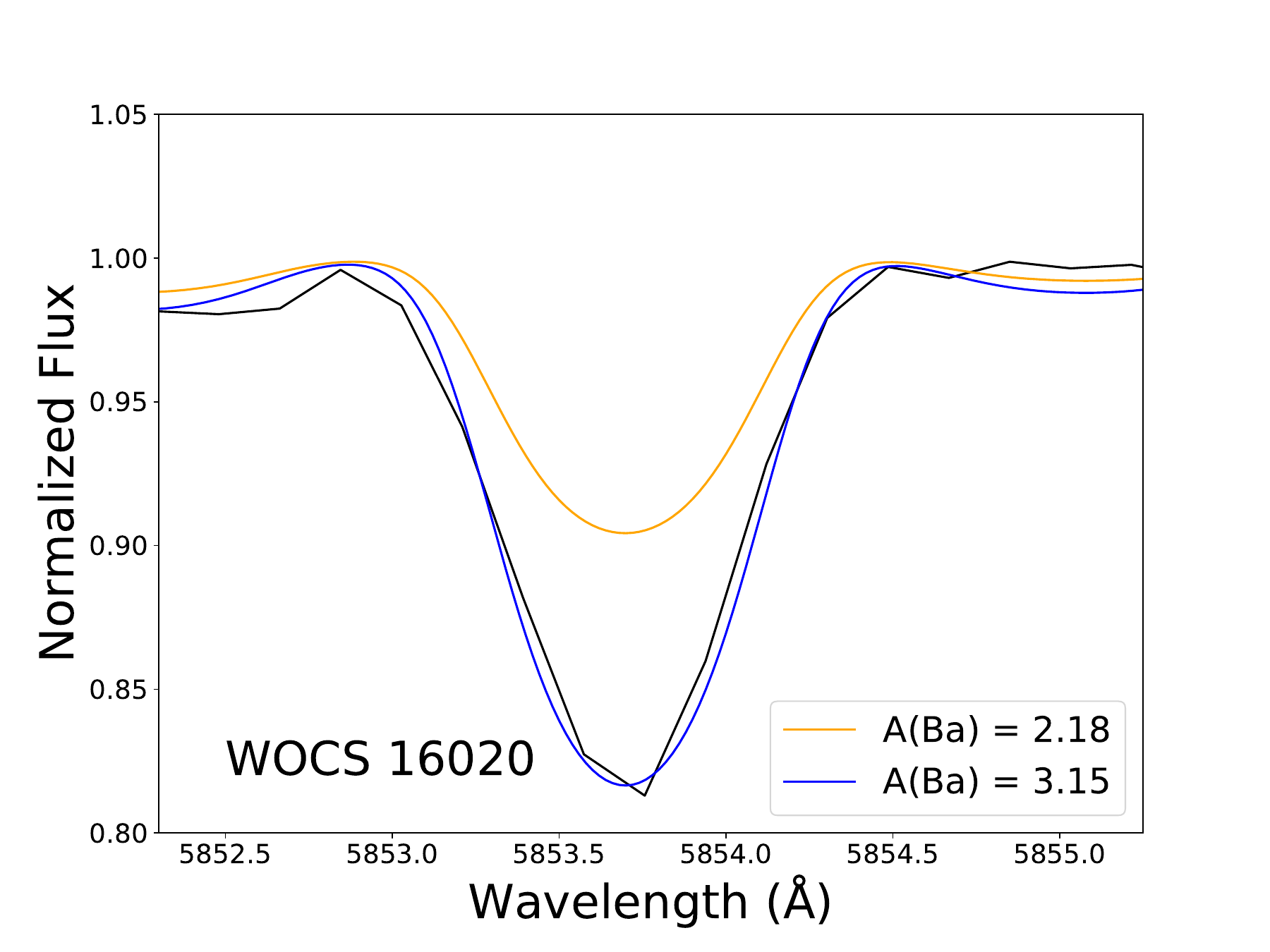}{0.4\linewidth}{}
  \fig{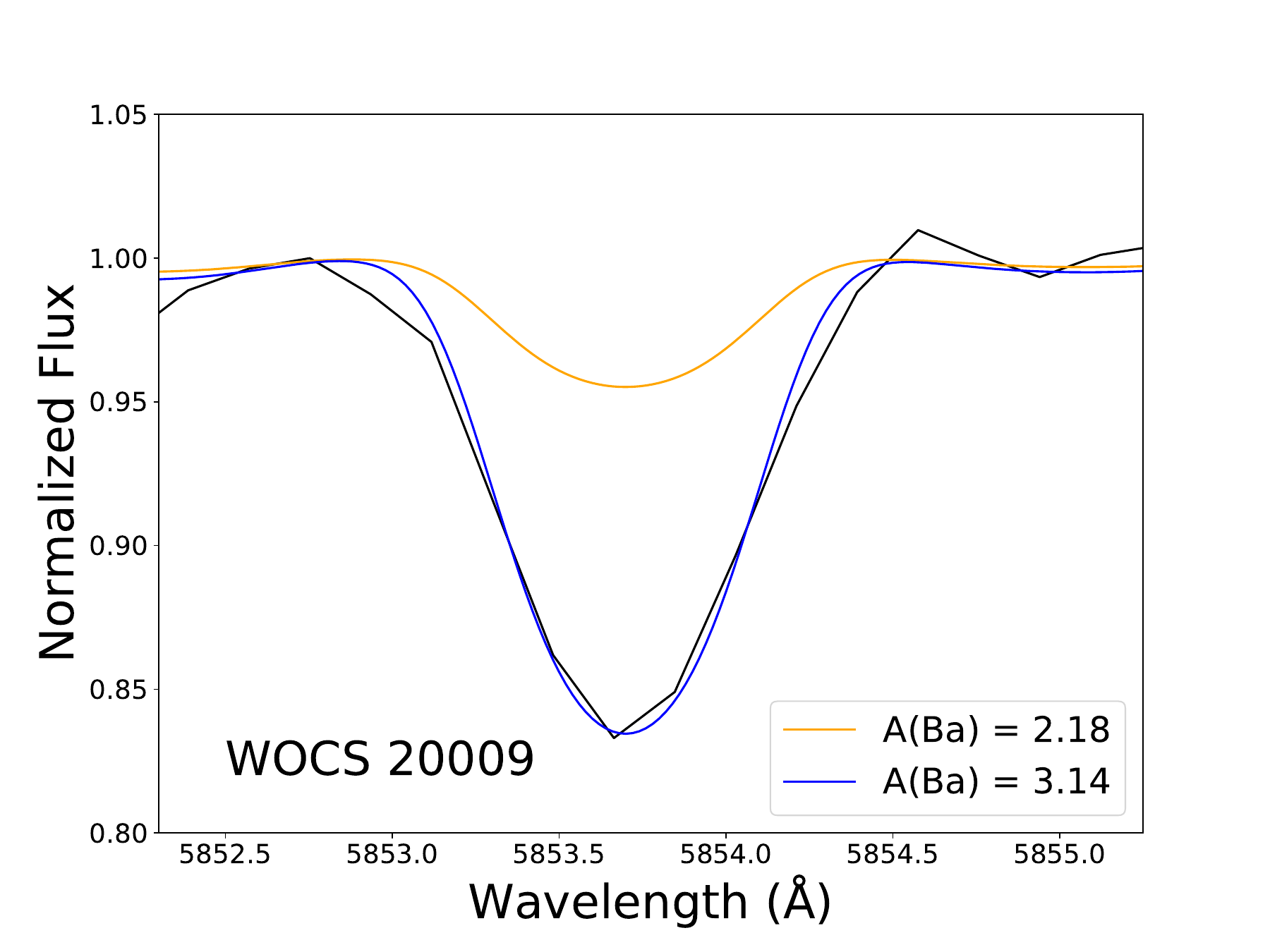}{0.4\linewidth}{}}
  \gridline{\fig{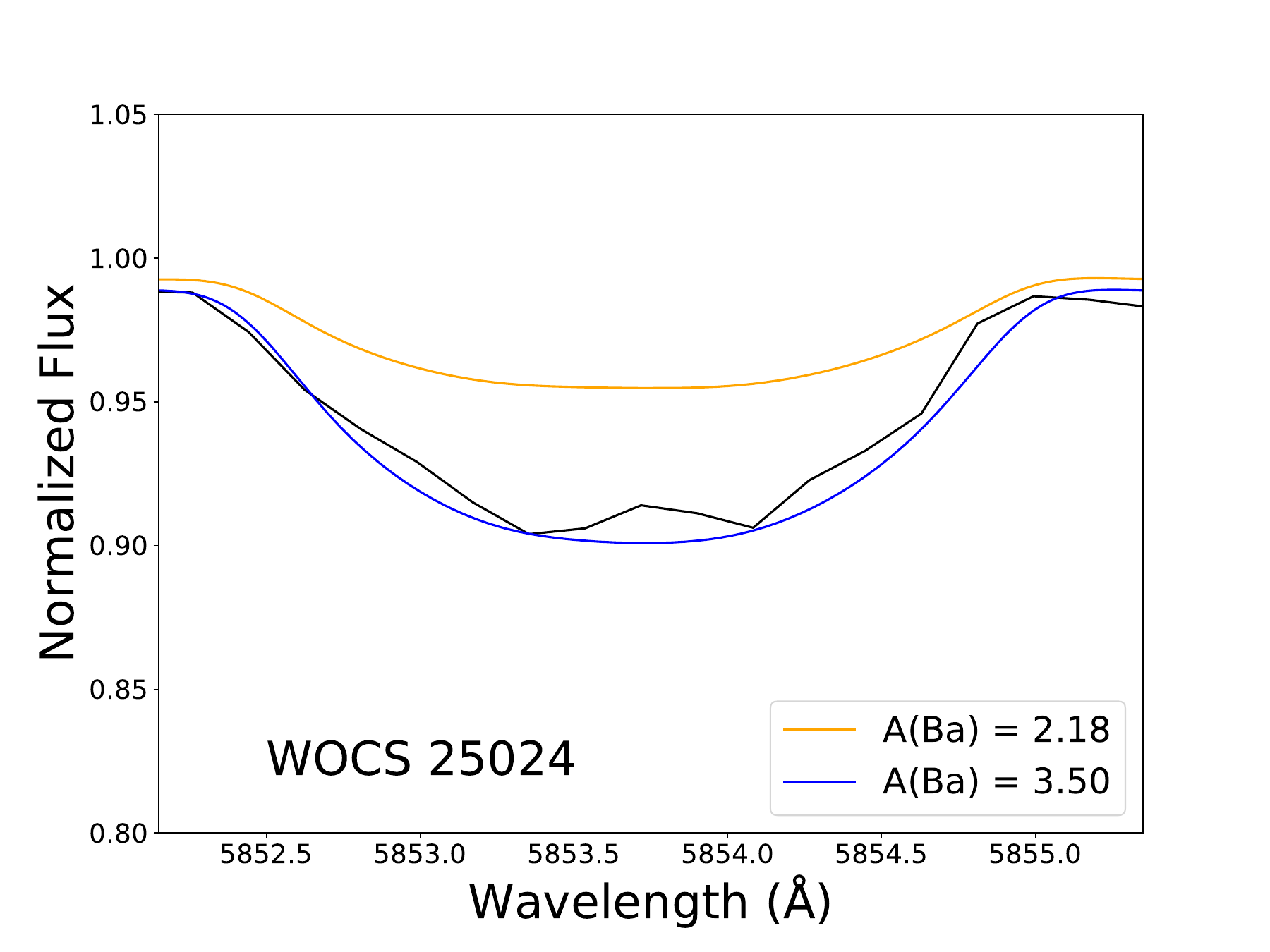}{0.4\linewidth}{}
  \fig{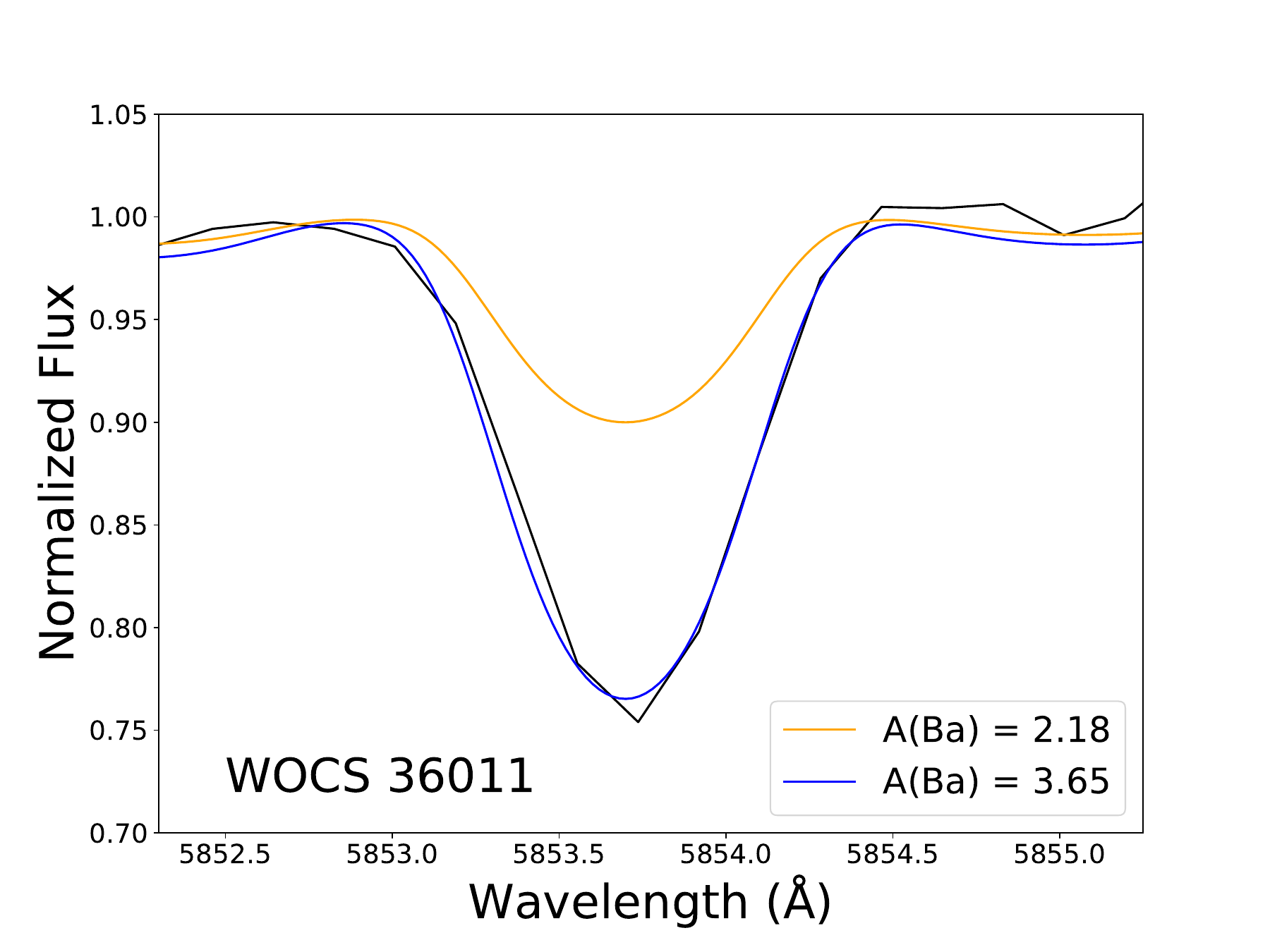}{0.4\linewidth}{}}
  \caption{Similar to Figure \ref{fig:m67_ba_specs}, but for the Ba-enriched BSSs of NGC 7789. The lines are the same as in Figure \ref{fig:m67_ba_specs}.}
  \label{fig:ngc7789_ba_specs}
\end{figure*}

As can be seen in Figure \ref{fig:bafe_plot}, there are six BSSs in NGC 7789 that are overabundant in Ba, i.e. above the 3$\sigma$ variation of the MS values, and which cannot be explained by varying assumptions regarding the selected stellar parameters: WOCS 5011, WOCS 10010, WOCS 16020, WOCS 20009, WOCS 25024, and WOCS 36011. In M67 there are four BSSs that are overabundant in Ba above the 3$\sigma$ level: WOCS 2011 (S968), WOCS 2013 (S1267), WOCS 3005 (S1263), and WOCS 3013 (S752). The Ba overabundances are also apparent in Figures \ref{fig:m67_ba_specs} and \ref{fig:ngc7789_ba_specs}, where we show the observed spectra compared to synthetic spectra.

\begin{deluxetable*}{c|ccccc|ccccc}
\tabletypesize{\scriptsize}
\tablecaption{Mean Cluster Abundances in NGC 7789 and M67}\label{tab:avg_abun}
\tablehead{\colhead{Ratio} & \multicolumn{5}{c}{\underline{NGC 7789}} & \multicolumn{5}{c}{\underline{M67}} \\
\colhead{} & \colhead{Our Work} & \colhead{\citetalias{Tautv2005}} & \colhead{\citetalias{Pancino2010}} & \colhead{\citetalias{Jacobson2011}} & \colhead{\citetalias{Overbeek2015}} & \colhead{Our Work} & \colhead{\citetalias{Tautv2000}} & \citetalias{Yong2005} & \colhead{\citetalias{Pancino2010}} & \colhead{\citetalias{Liu2019}}}
\tablewidth{0pt}
\startdata
$[$Fe/H$]$ & $-0.01 \pm 0.04$ & $-0.04 \pm 0.05$ & $+0.02 \pm 0.07$ & $+0.02 \pm 0.04$ & $+0.03 \pm 0.07$ & $-0.01 \pm 0.05$ & $-0.03\pm0.03$ & $+0.02\pm0.01$ & $+0.05\pm0.02$ & $-0.05\pm0.02$ \\
&&&&&&&&&&\\
$[$Ba/Fe$]$ & $-0.06 \pm 0.12$ & \nodata & $+0.47 \pm 0.05$ & \nodata & $+0.48 \pm 0.08$ & $+0.05 \pm 0.12$ & $+0.07\pm0.11$ & $-0.02\pm0.05$ & $+0.25\pm0.02$ & $+0.01\pm0.04$ \\
\enddata
\tablecomments{T00: \cite{Tautv2000}; T05: \cite{Tautv2005}; Y05: \cite{Yong2005}; P10: \cite{Pancino2010}; J11: \cite{Jacobson2011}; O15: \cite{Overbeek2015}; L19: \cite{Liu2019}.}
\end{deluxetable*}%
\centerwidetable
\begin{deluxetable*}{cccccc}
\tabletypesize{\footnotesize}
\tablecaption{M67 Barium Abundance Measurements\label{tab:ba_abuns}}
\tablehead{\colhead{WOCS ID} & $N$ & \colhead{[Fe/H]} & \colhead{Range}  & \colhead{[Ba/Fe]} & \colhead{Range}}
\tablewidth{0pt}
\startdata
\multicolumn{6}{l}{\textit{Blue Stragglers:}}\\
1006 & \nodata & $-$0.03 & $-$0.15 to +0.07 & 0.18 & +0.07 to +0.27 \\
1007 & \nodata & 0.02 & $-$0.10 to +0.10 & 0.10 & +0.04 to +0.19 \\
1025 & \nodata & $-$0.08 & $-$0.15 to +0.00 & 0.24 & +0.17 to +0.27 \\
1026 & \nodata & $-$0.06 & $-$0.20 to +0.05 & 0.02 & $-$0.13 to +0.12 \\
2009 & 7 & $-$0.30 & $-$0.43 to $-$0.20 & $-$0.30& $-$0.35 to $-$0.20 \\
2011 & 5 & $-$0.12 & $-$0.32 to +0.06 & 0.92 & +0.60 to +1.23 \\
2013 & 3 & 0.16 & +0.08 to +0.25 & 1.18 & +0.82 to +1.55 \\
3005 & 11 & 0.01 & $-$0.13 to +0.17 & 1.10 & +0.87 to +1.33 \\
3010 & 10 & $-$0.13 & $-$0.20 to $-$0.03 & 0.05 & $-$0.05 to +0.15 \\
3013 & \nodata & 0.04 & $-$0.05 to +0.15 & 1.09 & +0.78 to +1.46 \\
4006 & \nodata & $-$0.01 & $-$0.10 to +0.05 & 0.03 & +0.00 to +0.07 \\
5005 & 6 & $-$0.10 & $-$0.20 to +0.00 & $-$0.14 &  $-$0.18 to $-$0.08 \\
6038 & 9 & $-$0.12 & $-$0.19 to $-$0.11 & $-$0.23 & $-$0.30 to $-$0.14 \\
\\
\multicolumn{6}{l}{\textit{Blue Lurkers:}}\\
1020 & 9 & $-$0.08 & $-$0.17 to $-$0.05 & $-$0.22 & $-$0.29 to $-$0.16 \\
3001 & 7 & $-$0.06 & $-$0.17 to $-$0.01 & $-$0.08 & $-$0.24 to +0.09 \\
6025 & 9 & 0.02 & $-$0.11 to +0.07 & 0.01 & $-$0.11 to +0.14 \\
7035 & 11 &  $-$0.01 & $-$0.13 to +0.04 & 0.17 & +0.01 to +0.36 \\
9005 & 21 & $-$0.06 & $-$0.17 to $-$0.03 & 0.08 & $-$0.01 to +0.17 \\
12020 & 8 & $-$0.09 & $-$0.32 to +0.00 & 0.14 & +0.06 to +0.21 \\
14020 & 11 & 0.02 & $-$0.08 to +0.18 & 0.29 & +0.22 to +0.41 \\
\\
\multicolumn{6}{l}{\textit{Yellow Stragglers and Sub-subgiants:}}\\
1015 & 12 & 0.08 & $-$0.05 to +0.22 & 0.21 & +0.05 to +0.38 \\
2008 & 7 & 0.10 & $-$0.01 to +0.24 & 0.07 & $-$0.13 to +0.24 \\
13008 & 9 & $-$0.03 & $-$0.10 to +0.04 & 0.42 & +0.26 to +0.58 \\
\\
\multicolumn{6}{l}{\textit{Main Sequence:}}\\
3024 & 8 & $-$0.09 & $-$0.25 to $-$0.09 & $-$0.08 & $-$0.19 to $-$0.01 \\
5017 & 7 & $-$0.04 & $-$0.17 to +0.00 & 0.11 & $-$0.04 to +0.26 \\
5022 & 9 & $-$0.02 & $-$0.29 to +0.13 & 0.14 & $-$0.04 to +0.28 \\
7021 & 7 & $-$0.03 & $-$0.27 to +0.00 & 0.17 & $-$0.02 to +0.29 \\
7031 & 8 & $-$0.01 & $-$0.23 to +0.11 & 0.08 & $-$0.06 to +0.17 \\
7045 & 9 & 0.08 & $-$0.05 to +0.29 & $-$0.01 & $-$0.19 to +0.21 \\
8012 & 8 & $-$0.05 & $-$0.26 to +0.02 & 0.20 & +0.12 to +0.30 \\
8030 & 6 & 0.10 & $-$0.04 to +0.13 & 0.05 & $-$0.07 to +0.18 \\
10012 & 7 & $-$0.02 & $-$0.15 to +0.04 & 0.16 & +0.07 to +0.27 \\
10014 & 7 & $-$0.03 & $-$0.22 to +0.06 & $-$0.06 & $-$0.14 to +0.05 \\
11015 & 9 & $-$0.01 & $-$0.16 to +0.05 & $-$0.01 & $-$0.12 to +0.11 \\
11019 & 9 & 0.01 & $-$0.20 to +0.09 & 0.03 & $-$0.09 to +0.12 \\
12025 & 11 & 0.06 & $-$0.06 to +0.15 & 0.06 & $-$0.02 to +0.19 \\
12028 & 6 & 0.00 & $-$0.37 to +0.12 & $-$0.21 & $-$0.45 to +0.11 \\
16011 & 9 & $-$0.01 & $-$0.16 to +0.07 & $-$0.02 & $-$0.11 to +0.11 \\
\enddata
\end{deluxetable*}%
\centerwidetable
\begin{deluxetable*}{cccccc}
\tabletypesize{\footnotesize}
\tablecaption{NGC 7789 Barium Abundance Measurements\label{tab:ngc7789_ba_abuns}}
\tablehead{\colhead{WOCS ID} & $N$ & \colhead{[Fe/H]} & \colhead{Range}  & \colhead{[Ba/Fe]} & \colhead{Range}}
\tablewidth{0pt}
\startdata
\multicolumn{6}{l}{\textit{Blue Stragglers:}}\\
5004 & 5 & 0.07 & $-$0.11 to +0.28 & $-$0.02 & $-$0.10 to +0.12\\
5011 & \nodata & 0.03 & $-$0.10 to +0.20 & 1.32 & +0.82 to +1.87\\
10010 & \nodata & $-$0.10 & $-$0.29 to +0.09 & 0.88 & +0.80 to +0.96\\
10011 & 8 & $-$0.02 & $-$0.17 to +0.18 & 0.07 & $-$0.06 to +0.19\\
15015 & 5 & 0.06 & $-$0.07 to +0.20 & 0.27 & +0.17 to +0.42\\
16020 & 5 & $-$0.05 & $-$0.20 to +0.17 & 1.56 & +1.10 to +2.05\\
20009 & \nodata & 0.07 & $-$0.10 to +0.22 & 1.73 & +0.91 to +2.57\\
25008 & \nodata & 0.03 & $-$0.13 to +0.15 & 0.28 & +0.10 to +0.57\\
25024 & \nodata & 0.13 & $-$0.07 to +0.24 & 1.80 & +1.46 to +2.15\\
36011 & 6 & 0.16 & +0.00 to +0.35 & 1.77 & +1.35 to +2.20\\
\\
\multicolumn{6}{l}{\textit{Main Sequence:}}\\
5001 & 8 & 0.00 & $-$0.11 to +0.11 & $-$0.08 & $-$0.14 to $-$0.02\\
13015 & 5 & 0.03 & $-$0.05 to +0.15 & $-$0.11 & $-$0.20 to $-$0.05\\
14005 & 6 & $-$0.05 & $-$0.10 to +0.01 & 0.06 & +0.02 to +0.14\\
17016 & 5 & 0.04 & $-$0.09 to +0.16 & 0.16 & +0.07 to +0.26\\
18006 & 6 & 0.04 & $-$0.06 to +0.14 & $-$0.12 & $-$0.17 to $-$0.06\\
18007 & 5 & 0.00 & $-$0.10 to +0.08 & 0.08 & $-$0.02 to +0.20\\
23016 & 7 & $-$0.03 & $-$0.20 to +0.15 & 0.01 & $-$0.06 to +0.12\\
24008 & 5 & $-$0.05 & $-$0.19 to +0.08 & 0.08 & +0.01 to +0.18\\
25016 & 9 & $-$0.01 & $-$0.10 to +0.10 & $-$0.11 & $-$0.20 to $-$0.04\\
26009 & 3 & 0.00 & $-$0.10 to +0.10 & $-$0.27 & $-$0.35 to $-$0.22\\
27007 & 2 & $-$0.02 & $-$0.15 to +0.10 & $-$0.18 & $-$0.23 to $-$0.13\\
28015 & 5 & $-$0.02 & $-$0.20 to +0.15 & 0.01 & $-$0.06 to +0.06\\
33011 & 8 & $-$0.08 & $-$0.18 to +0.03 & $-$0.08 & $-$0.15 to $-$0.01\\
37011 & 3 & 0.03 & $-$0.10 to +0.15 & $-$0.10 & $-$0.18 to $-$0.03\\
37016 & 7 & $-$0.01 & $-$0.13 to +0.10 & $-$0.14 & $-$0.21 to $-$0.10\\
42011 & 8 & 0.00 & $-$0.14 to +0.13 & $-$0.24 & $-$0.28 to $-$0.16\\
42016 & 6 & $-$0.08 & $-$0.20 to +0.02 & $-$0.04 &  $-$0.08 to +0.02\\
\enddata
\end{deluxetable*}%
\vspace{-8em}
\section{Discussion}
\label{sec:discussion}

\subsection{Binarity of the Ba-enriched Blue Stragglers}
\label{subsec:binarity}

Mass transfer in binaries from an AGB primary star is thought to be responsible for Ba enrichment in both cluster stars and field stars. Longer orbital periods ($\gtrsim1000~\mathrm{d}$) are indicative of Case C mass transfer from an AGB companion (\citealt{Hurley2002, Chen2008}). We note that the WOCS RV period detection upper limit is 10$^4$ days (\citealt{Geller2012, Geller2021}), and that some Ba-enriched giants and dwarfs have been detected with orbital periods beyond this limit (see, e.g., \citealt{Fuhrmann2017, Escorza2023}). In particular, \cite{Escorza2023} identify two Ba dwarfs (HD 2454 and BD~$-$11$^{\circ}$3853) and one Ba giant (HD 119185), each of which have orbital periods $\gtrsim2\times10^4$~d. Substantial orbital eccentricities are also commonly observed among Ba giants and dwarfs (\citealt{Boffin2015, Jorissen2019, Escorza2019}).

The standard threshold for RV variability used in WOCS studies is that the ratio between the external error and the internal error $(e/i)$ is greater than 4 (\citealt{Geller2008}), above which we consider stars to be in binary systems. In our analysis we adopt the measured $e/i$ values from \cite{Geller2021} and \cite{Nine2020} for M67 and NGC 7789, respectively. Out of the four Ba-enriched BSSs in our M67 sample, two are velocity variable above this threshold: WOCS 2013 and WOCS 3013. Both of these BSSs have long orbital periods (Table \ref{tab:star_params}). WOCS 2013 has an orbital period of 846~d and an eccentricity of $e=0.48$, and WOCS 3013 has a period of 1003~d and $e=0.32$ (\citealt{Latham1996}). 

In NGC 7789, three Ba-enriched BSSs are velocity variable: WOCS 5011, WOCS 20009, and WOCS 36011 (\citealt{Nine2020}). Both WOCS 5011 and WOCS 20009 are known to be in long-period, moderately eccentric binaries. WOCS 5011 has an orbital period of 2710~d and an eccentricity of $e=0.32$, and WOCS 20009 has an orbital period of 4190~d and $e=0.27$. Previous observations of WOCS 36011 gave an orbital period of only 217.6~d and a very high eccentricity of $e=0.74$. Reanalysis including five more recent RV measurements suggests an even shorter-period ($\lesssim100$~d), eccentric binary orbit. The more recent RVs furthermore show long-term variability of 4~\kms\, over 4000~d, perhaps the result of a tertiary companion (E. Linck, private communication).

The comparatively short orbital period of the possible inner binary of WOCS 36011 in NGC 7789 is unusual among BSSs (\citealt{Mathieu2009}) and other mass-transfer products (see, e.g., \citealt{Oomen2020} and references therein). However, WOCS 36011 may be similar to the barium dwarf HD 48565 described by \cite{Escorza2019}. HD 48565 is a triple system in which the Ba-enriched primary star of the inner binary is in a short-period (73.3~d), eccentric ($e=0.22$) orbit with a much less luminous companion whose precise nature is as yet unknown. The outer triple companion of HD 48565 is in a very long-period ($>10,000$~d), eccentric ($e=0.4$) orbit, and is also much less luminous than the primary of the inner binary. \cite{Gao2023} proposed that this system may be the result of triple mass transfer, in which the outermost star in a hierarchical triple system evolves into an AGB star and transfers mass onto the inner binary, creating a Ba dwarf in the relatively short-period inner binary. WOCS 36011 may also be the product of triple mass transfer. 

Triple mass transfer may also be responsible for the creation of WOCS 4004 in NGC 6819, which is a double-lined spectroscopic binary ($P_{\mathrm{orb}}=297$~d, $e=0.227$) with a MS companion (\citealt{Milliman2014}), and in which only the primary star is enriched in Ba (\citetalias{Milliman2015}). The details of this process and of the final outcomes are sensitive to assumptions regarding three-body dynamics and non-conservative mass transfer (\citealt{Sepinsky2009, Toonen2016}). While we have not yet detected conclusive evidence of a triple companion in either NGC 7789 WOCS 36011 or NGC 6819 WOCS 4004, this possibility warrants future observational and theoretical study. 

The remaining BSSs enriched in Ba, WOCS 10010, WOCS 16020, and WOCS 25024 in NGC 7789 and WOCS 2011 and WOCS 3005 in M67, are not observed to be velocity variable. Similar Ba enhancements in non-velocity variable BSSs have been previously observed by \citetalias{Milliman2015} in NGC 6819, where four out of five Ba-enriched BSSs were not in detected binary systems. These non-velocity-variable BSSs constitute the majority of the Ba-enriched BSSs in the three open clusters; only $40\pm16$\% (6/15) of the Ba-enriched BSSs are observed to be in binaries with orbital periods $<10^4$~days. Current theories of Ba enrichment rely on the assumption that the material was accreted from a thermally-pulsing AGB companion (\citealt{BV1984, Webbink1988, Pols2003}), and observations of barium giants and dwarfs in the field have shown that the vast majority ($>85$\%) are in binary systems (\citealt{McClure1983, Jorissen1998, Escorza2019}).

The seeming contradiction of non-velocity-variable Ba-enriched BSSs may be resolved through either wind mass transfer (\citealt{Boffin1988, Theuns1993, Theuns1996}) or wind Roche-lobe overflow (WRLOF; \citealt{Mohamed2007, Mohamed2010}), in which chemically enriched material from the AGB donor star is accreted onto the secondary through stellar winds. Simulations of these forms of mass transfer conducted by \cite{BonMar2004} and \cite{Abate2013} predict the formation of Ba-enriched stars in binary systems with orbital periods up to of order 10$^5$ days, beyond the WOCS detection limit. Observations by \cite{Jorissen2019} revealed that a known S star, 57 Peg (HD 218634), is in a binary system with possible orbital periods of either $P_{\mathrm{orb}}\approx~4\times10^4$~d or $P_{\mathrm{orb}}\approx~2\times10^5$~d, within the range predicted by theory. It is therefore possible that the apparently single Ba-enriched BSSs in NGC 7789, NGC 6819, and M67 may in fact be in binary systems with similarly long orbital periods that we have not been able to detect through RV measurements.

\subsection{Amount of Barium Enhancement}
\label{subsec:amount}

The amount of barium enhancement among the Ba-rich BSSs in NGC 7789 ranges from [Ba/Fe]~$=+0.80$ (WOCS 10010) to [Ba/Fe]~$=+2.57$ (WOCS 20009). In M67, the Ba enrichment ranges from [Ba/Fe]~$=+0.60$ (WOCS 2011) to [Ba/Fe]~$=+1.55$ (WOCS 2013). The Ba abundances of WOCS 2011 (F153; \citealt{Fagerholm1906}) and WOCS 3005 (F185) in M67  as determined by \cite{Mathys1991}, who measured [Ba/Fe]~$=+0.61$ and [Ba/Fe]~$=+0.84$, respectively, are within our ranges of Ba abundances. The uncertainties in our measurements are much larger than those of \cite{Mathys1991} due to our inclusion of the sensitivity of Ba abundance measurements to assumptions in \vt\, (\citealt{Reddy2017, Spina2020}).

Following \citetalias{Milliman2015}, we compare our measurements of the Ba-enriched BSSs to the solar-metallicity AGB nucleosynthesis models available in the FUll-Network Repository of Updated Isotopic Tables \& Yields\footnote{\url{http://fruity.oa-teramo.inaf.it/modelli.pl}} (F.R.U.I.T.Y.; \citealt{Cristallo2016} and references therein). The nucleosynthetic yields of the 1.5, 2.0, and 2.5~\Msolar\, solar-metallicity ($Z=0.014$) F.R.U.I.T.Y. models at their last third dredge-up event predict a final surface Ba abundance for their 1.5 \Msolar\; model of [Ba/Fe]~$+0.65$; for their 2.0~\Msolar\; model, the final Ba yield is [Ba/Fe]~$=+1.03$; and for their 2.5~\Msolar\; model, the final Ba yield is [Ba/Fe]~$=+1.13$. All final yields are calculated assuming the standard $^{13}$C pocket model (see \citealt{Cristallo2015}). 

Note that the uncertainties in measured Ba enhancements in the BSSs of NGC 7789 and M67 are such that we cannot constrain precise AGB donor masses on an individual basis. Furthermore, the observed Ba enrichments of the BSSs may not be reflective of the initial Ba abundance deposited on the accretor star. When enriched material is deposited onto the surface of the accreting star, it is subject to various and competing mixing processes, including diffusion (\citealt{Michaud1970, Proffitt1989, Deal2020}), radiative acceleration (\citealt{Vick2010, Xiang2020}), thermohaline mixing (\citealt{Vauclair2004, Brown2013}), gravitational settling (\citealt{Vauclair1982, Thompson2008}), and mixing due to rotation (\citealt{Talon2006, Quievy2009}). The current observed surface abundances may therefore be altered from what was initially deposited on the surface of the accreting star.

Broadly, our results are consistent with AGB stars more massive than about 1.5~\Msolar\ being responsible for the creation of the Ba-enriched BSSs in NGC 7789 and M67. This mass range is consistent with the turnoff masses of these clusters. The range of Ba enhancements in M67 ([Ba/Fe]~=~+0.60 to +1.55) permits initial AGB donor masses of 1.5~\Msolar\, for which theory predicts final Ba yields of [Ba/Fe]~=~+0.65~dex. The range of Ba enhancements in NGC 7789 ([Ba/Fe]~=~+0.80 to +2.57) permits AGB donor masses of approximately 2.0~\Msolar, for which a final Ba yield of [Ba/Fe]~=~+1.03 is predicted (\citealt{Cristallo2016}).

The absence of Ba enrichment in the remaining BSSs does not by itself rule out formation through AGB mass transfer, as mass transfer may have occurred after helium core exhaustion and before the donor star reached the thermally-pulsing phase. In such cases, carbon abundances may still be telling. Empirical evidence to date, however, is unclear on the sense of variation in carbon abundances among BSSs. In some cases there may be enhancement in the carbon abundance of the accreting star in a similar manner as the carbon-enhanced metal-poor stars (\citealt{Aoki2002, Beers2005}). Recent observations by \cite{Brady2023} found evidence for possible carbon enhancements in WOCS 1025 and WOCS 3010 in M67, neither of which are Ba-enhanced but both of which are in long-period orbits of order 1000 days (see Table \ref{tab:star_params}). However, AGB mass transfer in other cases suggests carbon depletion (\citealt{Sarna1996, Ferraro2006}). This was observed by \cite{Mathys1991} in M67, who measured [C/Fe]~=~$-0.75$ in WOCS 2011 and [C/Fe]~=~$-0.64$ in WOCS 3005. \cite{Shetrone2000} and \cite{BM2018} found that the remaining BSSs in M67 have carbon abundances consistent with the MS. No such detailed abundance study has been performed for the BSSs of NGC 7789, though this is the subject of future work.

The two yellow stragglers we observed in M67 do not show evidence of Ba enrichment, consistent with the observations of these stars by \cite{McGahee2014}. The blue lurkers of M67 also do not show evidence of Ba enrichment, which may be expected if their predominant formation mechanism is mass transfer from an RGB star, as hypothesized by \cite{Nine2023}. The hypothesis of RGB mass transfer is supported by the frequency of shorter-period orbits ($P_{\mathrm{orb}}\sim100$~d) among the blue lurkers, indicative of Case B mass transfer from an evolved star prior to the helium flash (\citealt{Plavec1968, Ziolkowski1970}). Finally, the sub-subgiant WOCS 13008 (S1063) shows some evidence of Ba enrichment, with a range of 1.8$\sigma$ to 4.4$\sigma$ in possible Ba enrichment compared to the MS. WOCS 13008 is known to be magnetically active and heavily spotted (\citealt{Geller2017, Gosnell2022, Leiner2022}), which makes our assumptions of \vt\, for this star overly simplistic and our formal uncertainties likely underestimated. For this reason we do not consider WOCS 13008 to be securely Ba-enriched.

\subsection{Comparison with Field Barium Dwarfs}
\label{subsec:field_dwarfs}

\begin{figure}[t!]
    \centering
    \includegraphics[width=\linewidth]{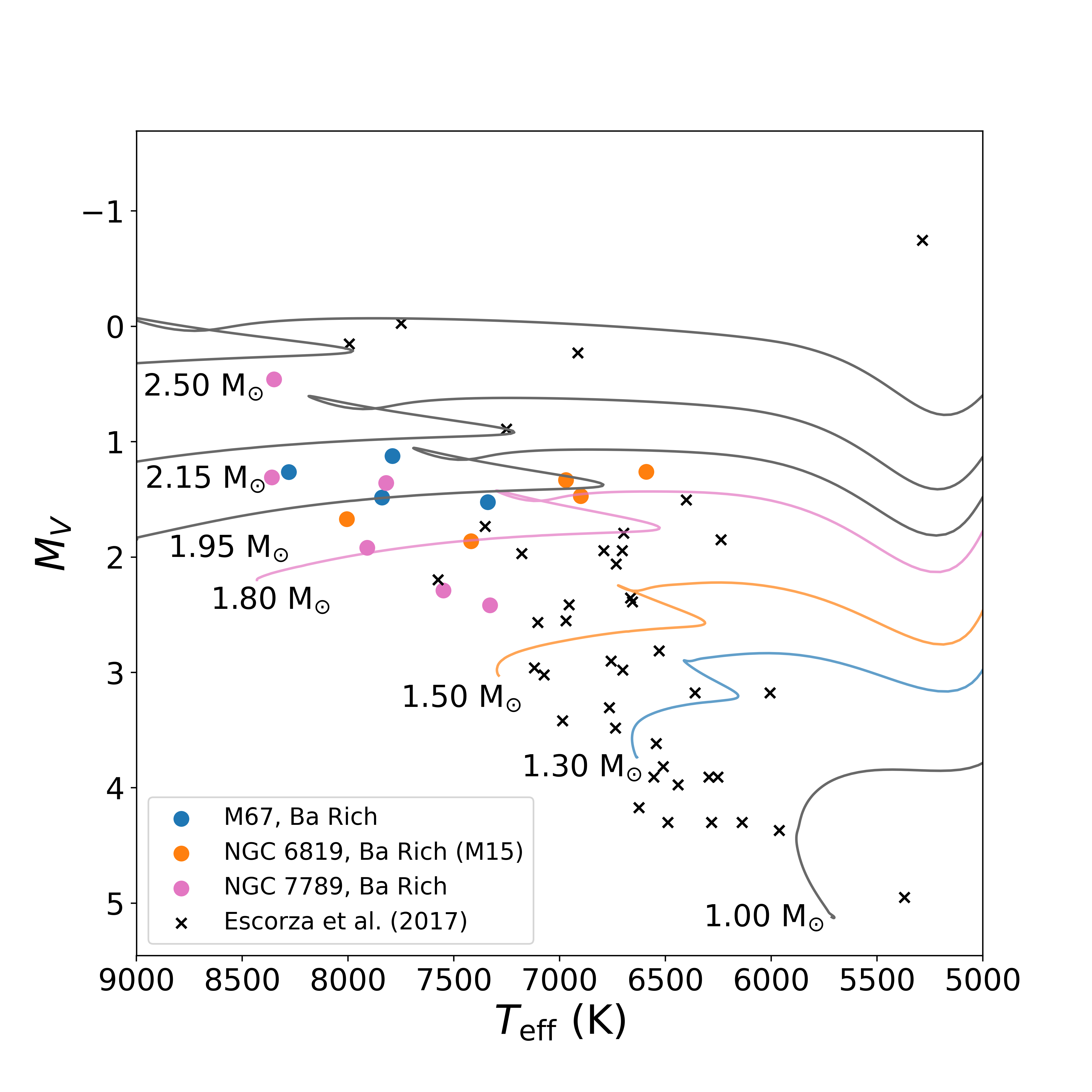}
    \caption{HR diagram of the Ba-enriched BSSs (colored points) compared to the field Ba dwarfs (black ``x'' marks) from \cite{Escorza2017}. We highlight the turnoff masses of NGC 7789, NGC 6819, and M67 with MIST single-star, solar-metallicity evolutionary tracks with the same color as the Ba-enriched BSSs of the respective clusters. We also plot other selected solar-metallicity MIST evolutionary tracks for context.}
    \label{fig:hr_diagram_escorza}
\end{figure}

In Figure \ref{fig:hr_diagram_escorza}, we compare the Ba-enriched BSSs of NGC 7789, NGC 6819, and M67 to the sample of field Ba dwarfs from \citet[see also \citealt{Escorza2019}]{Escorza2017} on the HR diagram. We compute the absolute $V$ magnitude of the Ba-enriched BSSs using cluster distances from \cite{Hunt2023}, as well as apparent $V$ magnitudes drawn from \cite{Gim1998} for NGC 7789, \cite{Yang2013} for NGC 6819, and \cite{Montgomery1993} for M67. We furthermore adopt values of $E(B-V)$ from \cite{Wu2007} for NGC 7789, \cite{Yang2013} for NGC 6819, and \cite{Taylor2007} for M67. We also plot MIST single-stellar, solar-metallicity evolutionary tracks in order to give an estimate of the approximate range in BSS masses. 

The Ba-enriched BSSs of all three open clusters occupy a strikingly compact region of the HR diagram, between approximately 1.6~\Msolar\, and 2.5~\Msolar\, based on the MIST evolutionary tracks. The BSSs occupy a region of the HR diagram consistently brighter and hotter, and therefore more massive, than the field Ba dwarf distribution. This may be the result of an observational bias selecting younger and consequently more massive stars in open clusters compared to the field, as discussed in \cite{Escorza2019}. The BSSs also occupy the same region of the HR diagram regardless of the turnoff mass of their respective host clusters, which we highlight in Figure \ref{fig:hr_diagram_escorza} with colored MIST evolutionary tracks. We discuss this finding further in Section \ref{subsec:ba_cluster_age}. 

\cite{Escorza2019} found a peak in the field Ba dwarf mass distribution at approximately 1.2~\Msolar, while our Ba-enriched BSSs are more massive than this peak. On the other hand, the isochronal masses of the Ba-enriched BSSs in these old open clusters (Figure \ref{fig:hr_diagram_escorza}) are similar to the field Ba giant mass distribution observed by \cite{Jorissen2019}, who determined a peak at $\sim2.0$ \Msolar. The Ba-enriched BSSs in these open clusters may therefore be similar to the progenitors of the field Ba giants observed by \cite{Jorissen2019}. It is possible that these open clusters may also host Ba-enriched dwarf stars of similar masses as those found by \cite{Escorza2019}. They would not appear as classical BSSs, but rather as blue lurkers since they would not have gained enough mass to appear above the MSTOs of these clusters. Our selection criterion of non-rapidly-rotating MS control stars precludes the detection of Ba-enriched blue lurkers in these clusters; such a possibility warrants future study.

\subsection{Blue Straggler Barium Enrichment as a Function of Cluster Age}
\label{subsec:ba_cluster_age}

\begin{figure*}[t!]
    \includegraphics[width=\linewidth]{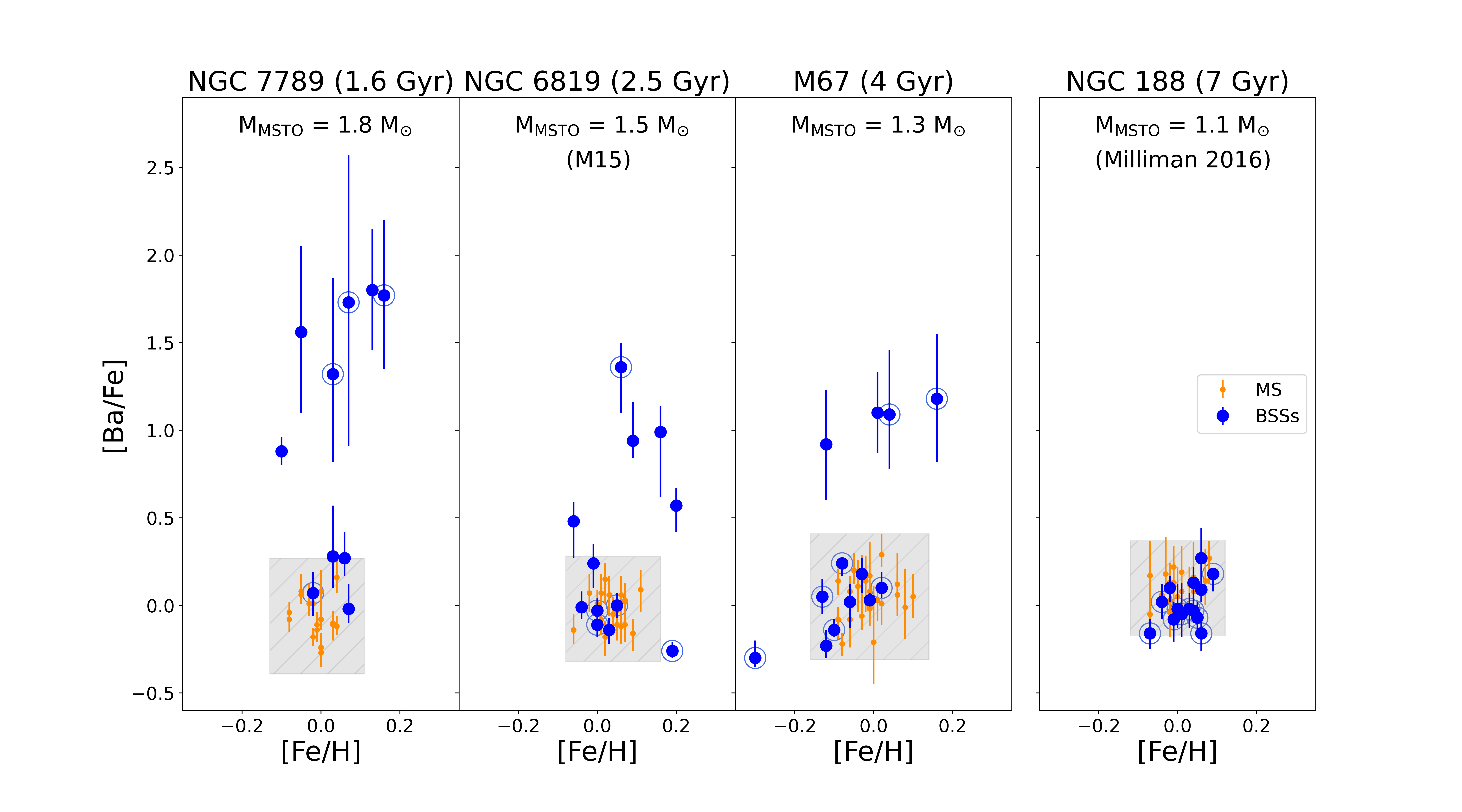}
    \caption{[Ba/Fe] vs. [Fe/H] diagrams of the four clusters. The points and boxes are the same as in Figure \ref{fig:bafe_plot}. The NGC 188 [Ba/Fe] diagram is plotted separately, as theory predicts that AGB stars at this turnoff mass do not produce Ba (Section \ref{subsec:ba_cluster_age}). We also include for reference the turnoff mass of each cluster in their respective plot.}
    \label{fig:multicluster_bafe_plots}
\end{figure*}

\begin{figure*}[t!]
    \includegraphics[width=\linewidth]{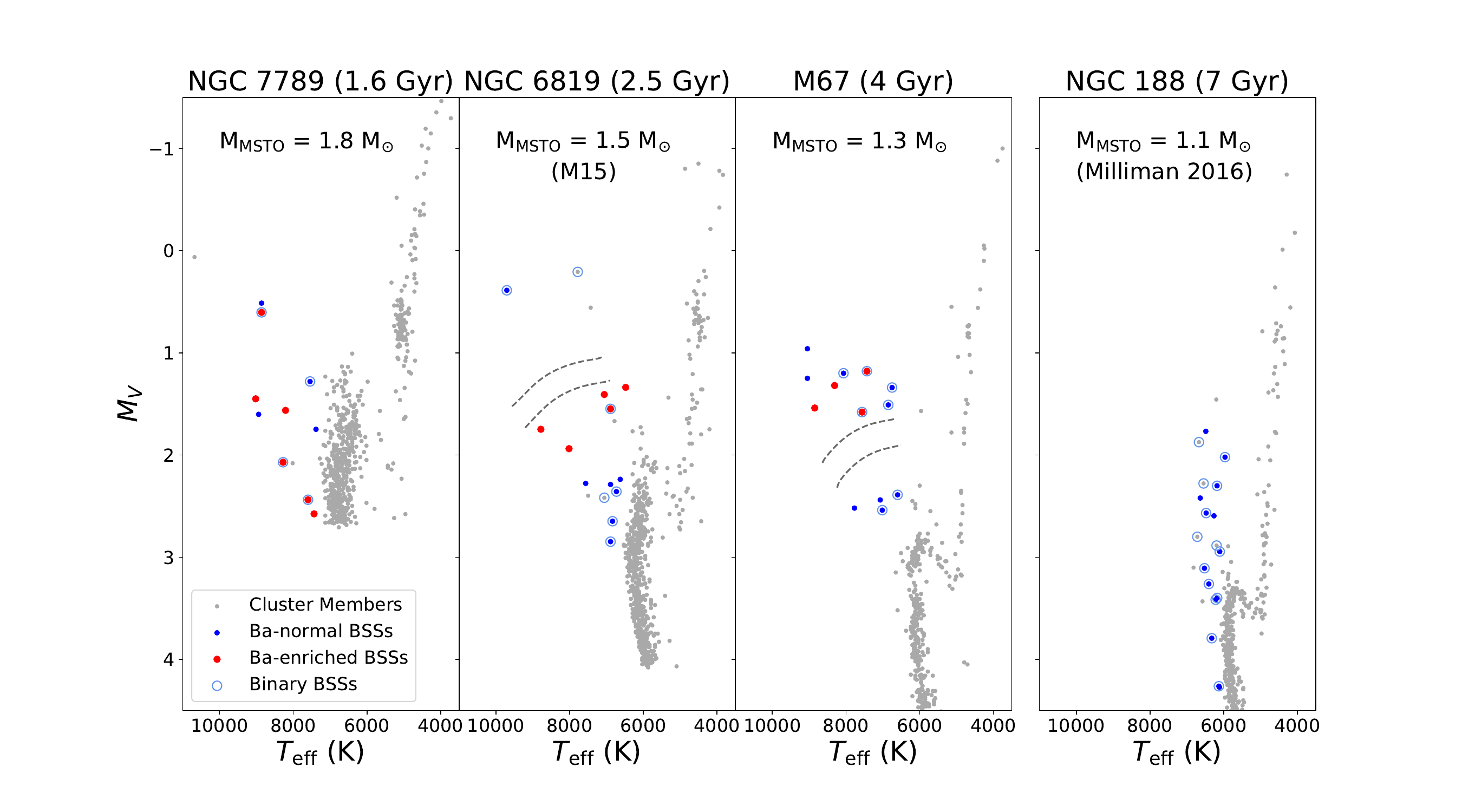}
    \caption{HR diagrams of the four open clusters considered in this work. We indicate the BSSs observed for Ba abundances with blue points, and the BSSs that are enhanced in Ba compared to the MS are highlighted in red. BSSs in known binary systems are indicated with a blue circle. We also show the gap in BSS mass distribution in intermediate-age open clusters (2 to 4 Gyr) found by \cite{Leiner2021}, centered at 0.5 \Msolar\, above the MSTO of NGC 6819 and M67. We again plot the HR diagram for NGC 188 separately, and include the turnoff mass of each cluster as in Figure \ref{fig:multicluster_bafe_plots}.}
    \label{fig:multicluster_plots}
\end{figure*}

The degree of Ba enrichment in the BSSs is likely related to the mass of the AGB donor star, in that more massive AGB stars (up to $\sim2.5$--$3.0$ \Msolar) bring greater amounts of Ba to their surfaces through third dredge-up events (\citealt{Iben1983, Lugaro2012, Cristallo2016}). In Section \ref{subsec:field_dwarfs}, we find that the Ba-enriched BSSs occupy similar regions of the HR diagram regardless of cluster age (Figure \ref{fig:multicluster_plots}). Here we also find an anticorrelation between the amount of Ba enrichment in the BSSs and cluster age, indicative of mass transfer from less massive AGB stars in older open clusters.  

Incorporating the Ba abundance results from \citetalias{Milliman2015} for NGC 6819 and \cite{Milliman2016} for NGC 188, we can explore the trend of Ba enrichment across cluster age. In Figure \ref{fig:multicluster_bafe_plots} we show in age sequence the [Ba/Fe] vs. [Fe/H] diagrams of the BSSs and MS samples of the four open clusters. There is a trend of decreasing fractions of BSSs that are Ba-enriched, with 60$\pm$24\% (6/10) being enriched in NGC 7789, 42$\pm$19\% (5/14) in NGC 6819, and 31$\pm$15\% (4/13) in M67. This trend may reflect that, as a cluster ages and AGB stars get closer to the lower limit of 1.3~\Msolar\, for Ba enrichment through third dredge-up, there are fewer binaries available to produce Ba-enriched BSSs.

There is also a trend of decreasing [Ba/Fe] enrichment among the BSSs of the four open clusters, from a typical enrichment of [Ba/Fe]~$\sim$~1.5 dex in NGC 7789 (1.6 Gyr) to an absence of Ba enrichment in NGC 188 (7 Gyr). This trend is consistent with models of AGB nucleosynthesis which predict decreasing Ba yields with decreasing AGB mass, and that solar-metallicity AGB stars less massive than about 1.3 \Msolar, do not undergo third dredge-up and therefore do not bring Ba up to their surfaces (\citealt{Busso1999, Karakas2016}). The MSTO of NGC 188 is 1.1 \Msolar.

In Figure \ref{fig:multicluster_plots} we show the HR diagrams of the four open clusters in sequence of age. We computed the absolute $V$ magnitudes as described in Section \ref{subsec:field_dwarfs}, including also the apparent magnitudes and $E(B-V)$ for NGC 188 as determined by \cite{Sarajedini1999}. In order to estimate the effective temperatures for all cluster member stars, we dereddened their measured colors according to the relations from \cite{Cardelli1989} and computed their \teff\, using the prescription of \cite{Ramirez2005}. We highlight the Ba-enriched BSSs with red points. The locations of the Ba-enriched BSSs on the HR diagrams are independent of the locations of the MSTOs or MSs of their host clusters. We include for reference the current turnoff mass of each cluster.

The locations of the Ba-enriched BSSs on the HR diagrams in these two clusters suggest that the AGB mass transfer was least efficient in NGC 7789 and most efficient in M67. This may also indicate a change in the dominant mode of AGB mass transfer, for example from wind mass transfer to WRLOF or classical Case C Roche-lobe overflow (e.g., \citealt{Sun2021, Sun2023}).

In Figure \ref{fig:multicluster_plots} we also present for comparison the location of a gap in BSS mass distributions of intermediate-age open clusters such as NGC 6819 and M67 as found by \cite{Leiner2021}, which they determine to be centered at 0.5~\Msolar\, above the MSTOs of such clusters. These authors hypothesize that this gap appears as a consequence of different efficiencies of mass transfer, with less efficient mass transfer creating BSSs below the mass gap and more efficient mass transfer creating those above. This trend also may be reflected in the Ba-enrichment of the BSSs of NGC 6819 and M67. The Ba-enriched BSSs of NGC 6819 all lie below the mass gap found by \cite{Leiner2021}. In addition, four out of five of the Ba-enriched BSSs are not velocity variable, suggestive of very wide binaries with comparatively inefficient mass transfer. In M67, however, all of the Ba-enriched BSSs lie above the BSS mass gap, and two of the four Ba-enriched BSSs are in closer binaries, permitting more conservative modes of mass transfer.

This trend in increasing mass transfer efficiency with cluster age might be understood through considering, in the case of classical Roche-lobe overflow or WRLOF, the response of the donor to mass loss ($\zeta_{\mathrm{ad}}$) to the response of the Roche lobe radius to mass loss ($
\zeta_{\mathrm{RL}}$). From \cite{Woods2012}, these quantities are defined to be:

\begin{equation}
    \zeta_{\mathrm{ad}} \equiv \left(\frac{\partial\log R_d}{\partial\log M_d}\right)_{\mathrm{ad}},
\end{equation}

\begin{equation}
    \zeta_{\mathrm{RL}} \equiv \frac{\partial\log R_{\mathrm{RL}}}{\partial\log M_d},
\end{equation}

\noindent where $M_d$ and $R_d$ are the mass and radius of the donor star, respectively. Mass transfer is thought to be stable as long as it satisfies the condition (\citealt{Soberman1997, Ivanova2015, Sun2021}):

\begin{equation}
\label{eqn:criterion}
    \zeta_{\mathrm{RL}} \leq \zeta_{\mathrm{ad}}.
\end{equation}

Following \cite{Soberman1997}, assuming that the structure of the donor star can be well approximated by a condensed polytrope as described in \cite{Hjellming1987}, the adiabatic response of the donor to mass loss depends on the core mass fraction of the donor star such that

\begin{equation}
\label{eqn:soberman}
\begin{split}
    \zeta_{\mathrm{ad}} = \frac{2}{3}\frac{m}{1-m} - \frac{1}{3}\frac{1-m}{1+2m} - 0.03m \\ + 0.2\left(\frac{m}{1+(1-m)^{-6}}\right),
\end{split}
\end{equation}

\noindent where $m=M_{\mathrm{core}}/M_{\mathrm{tot}}$.

From \cite{Woods2012}, the Roche lobe response can be expanded as:

\begin{equation}
\label{eqn:zeta_roche}
    \zeta_{\mathrm{RL}} = \frac{\partial\ln a}{\partial\ln M_d} + \frac{\partial\ln (R_{\mathrm{RL}}/a)}{\partial\ln q}\frac{\partial\ln q}{\partial\ln M_d},
\end{equation}

\noindent where $q$ is the mass ratio $\left(M_{\mathrm{donor}}/M_{\mathrm{accretor}}\right)$. 
The first term in the expansion is the change in orbital separation due to mass loss from the donor star. Assuming that the mass loss rate from the donor star due to mass transfer is much greater than mass loss due to winds, the first term can be written in terms of the mass ratio $q$ (\citealt{Ivanova2015}):

\begin{equation}
    \frac{\partial\ln a}{\partial\ln M_d} = \frac{2q^2 -2 - q(1-\beta)}{q+1},
\end{equation}

\noindent where $\beta$ is the mass transfer efficiency ($\beta=1$ for fully conservative mass transfer, $\beta=0$ for fully non-conservative mass transfer). The first factor of the second term of the expansion is the change in the relative Roche lobe radius of the donor with respect to the change in mass ratio, and can be written as (\citealt{Eggleton1983, Soberman1997}):

\begin{equation}
    \frac{\partial\ln (R_{\mathrm{RL}}/a)}{\partial\ln q} = \frac{2}{3} - \frac{q^{1/3}}{3}\frac{1.2q^{1/3} + 1/\left(1+q^{1/3}\right)}{0.6q^{2/3}+\ln\left(1+q^{1/3}\right)}.
\end{equation}

\noindent The second factor of the second term is the change in mass ratio due to mass loss from the system:

\begin{equation}
    \frac{\partial\ln q}{\partial\ln M_d} = 1 + \beta q.
\end{equation}

\noindent For a more detailed analysis of the physics of non-conservative mass transfer in the creation of BSSs, we refer the reader to \cite{Sun2021}.

\begin{figure*}[htbp]
    \gridline{\fig{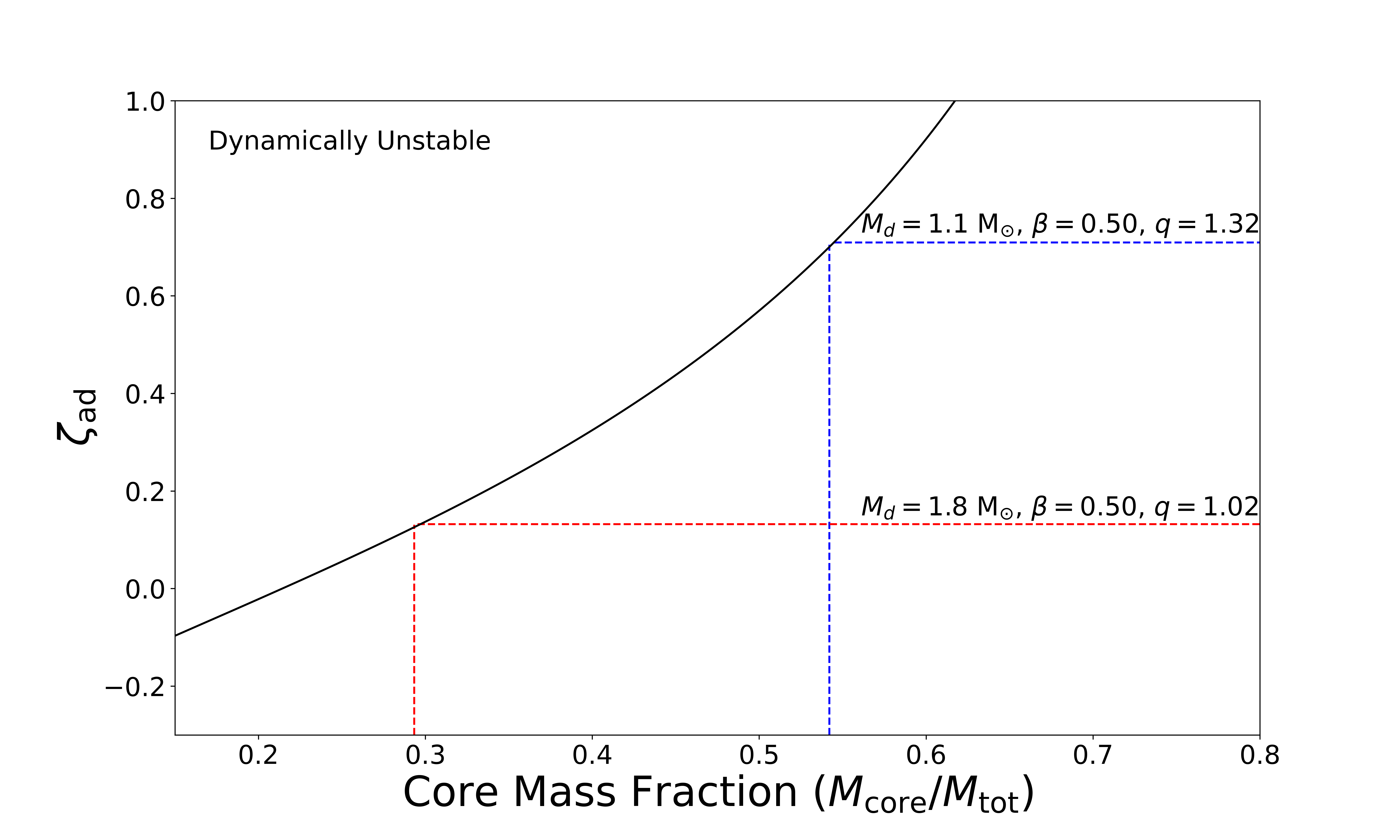}{0.5\linewidth}{}
    \fig{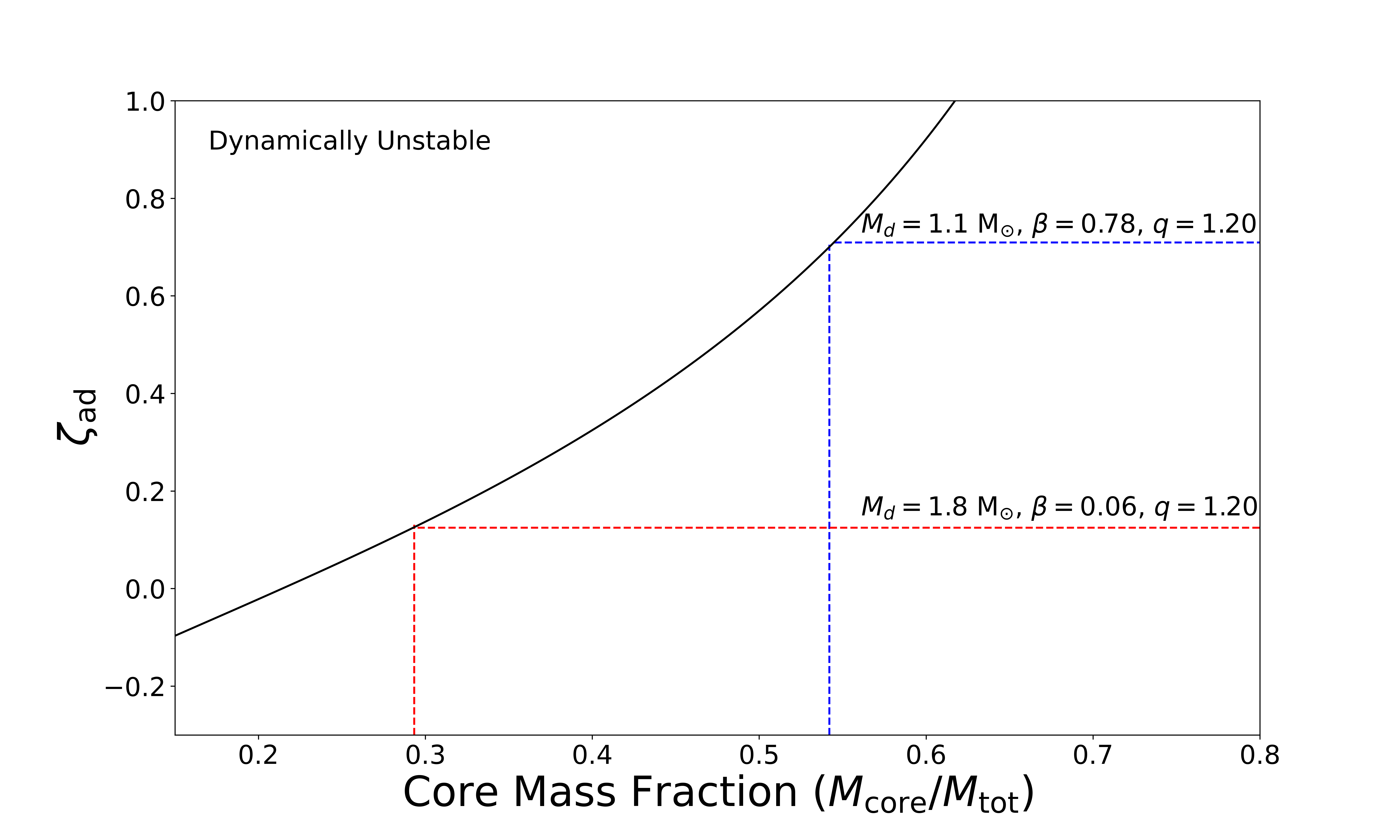}{0.5\linewidth}{}}
    \caption{Mass transfer stability plots, considering initial donor masses of 1.1~\Msolar\, (blue dashed lines) and 1.8~\Msolar\, (red dashed lines). On the left-hand plot, we consider mass transfer assuming the same mass transfer efficiency ($\beta=0.5$), and on the right we assume the same initial mass ratio ($q=1.20$). The black lines in both plots trace the adiabatic response of the donor to mass loss ($\zeta_{\mathrm{ad}}$; see Equation \ref{eqn:soberman}) as a function of core mass. The vertical lines mark the core mass ratios of a 1.1~\Msolar\, (blue) and 1.8~\Msolar\, (red) AGB star from our MESA models. The horizontal lines mark the corresponding response of the Roche lobe to mass loss ($
    \zeta_{\mathrm{RL}}$; see Equation \ref{eqn:zeta_roche}) in both cases. The regions to the left of the black lines are dynamically unstable to mass transfer ($\zeta_{\mathrm{RL}} > \zeta_{\mathrm{ad}}$). Compare with Figure 1 of \cite{Woods2012} or Figure 2 of \cite{Ivanova2015}.}
    \label{fig:mt_stability}
\end{figure*}

In Figure \ref{fig:mt_stability} we consider the stability of mass transfer from a 1.1~\Msolar\, AGB donor star, the current turnoff mass of NGC 188, and from a 1.8~\Msolar\, AGB star, the turnoff mass of NGC 7789, under the condition imposed by Equation \ref{eqn:criterion}. We calculate the core mass ratios of both stars from simulations of single-stellar evolution computed with the \texttt{Modules for Experiments in Stellar Evolution}\footnote{\url{https://docs.mesastar.org/en/latest/index.html}} (MESA, version r15140\footnote{Compiled with MESA SDK version 21.11.1 for Mac OS (\citealt{Townsend2021})}; \citealt{Paxton2011, Paxton2013, Paxton2015, Paxton2018, Paxton2019}). Our inlists are publicly available at \url{https://zenodo.org/records/11176102}. We consider mass transfer from these stars under two sets of circumstances: the first under the assumption of identical mass transfer efficiencies (which we select to be $\beta=0.5$), and the second under the assumption of identical initial mass ratios (which we select to be $q=1.20$). 

We find that under the assumption of equal mass transfer efficiencies, the maximum mass ratio for stable mass transfer for a 1.1~\Msolar\, AGB donor is $q\approx1.32$, while for a 1.8~\Msolar\, donor the maximum mass ratio for stable mass transfer is only $q\approx1.02$. Under the assumption that the secondary mass distribution of binaries that are sufficiently widely separated to undergo Case C mass transfer is approximately uniform (\citealt{Raghavan2010, Moe2017}), this implies that it is more probable for a lower-mass AGB donor to undergo stable mass transfer at a given efficiency. Similarly, under the assumption of equal initial mass ratios we find that the maximum mass transfer efficiency for stable mass transfer from a 1.1~\Msolar\, donor is $\beta\approx0.78$, while for a 1.8~\Msolar\, donor the maximum stable efficiency is only $\beta\approx0.06$. This implies that mass transfer through classical Roche-lobe overflow or WRLOF at a given mass ratio must be less conservative for more massive AGB donors in order for it to be stable. 

These results put together indicate that more massive AGB donors tend to experience less conservative mass transfer than less massive AGB donors, and thereby transfer proportionally less mass compared to the MSTO of their host clusters. The Ba-enriched BSSs of M67 received mass from their thermally-pulsing AGB donors through more conservative mass transfer than those in NGC 7789, which would place the M67 Ba-enriched BSSs brighter in magnitude with respect to their MSTO than those in NGC 7789, as observed in Figure \ref{fig:multicluster_plots}. 

Further, the interplay of the initial mass of the donor star and the resulting mass transfer efficiency may explain why the Ba-enriched BSSs occupy the same region of the HR diagram regardless of the age and turnoff masses of their host clusters. Although mass transfer from a higher-mass AGB star in NGC 7789 would tend to be less efficient, the turnoff stars are more massive, and the donor star has $\sim$1.2~\Msolar\, in its envelope available to donate, based on our MESA models. For NGC 188, the mass transfer efficiency is more efficient, but the turnoff stars are less massive, and the donor star also has less material available to donate ($\sim$0.4~\Msolar, based on our MESA models). The combined effects may explain the similar luminosities of Ba-enriched BSSs on the HR diagrams of older open clusters. While it is not forbidden for a higher-mass AGB star to undergo unstable mass transfer either through a high mass ratio or high mass transfer efficiency, the result of that process may not appear as a classical BSS.

This analysis makes the assumption that the structure of the donor star can be described by a condensed polytrope, which may not be accurate for AGB stars due to radiation pressure in the envelope (\citealt{Hjellming1987}). More recent studies of mass transfer stability, however, have shown that while mass transfer may be more stable and more conservative at a given core mass fraction than this analysis implies, it is still true that a less massive evolved donor is more likely to undergo more conservative mass transfer than a higher-mass donor (\citealt{Pavlovskii2015, Ge2020, Temmink2023}).

In summary, there is a suggestion in these data of a connection between the initial mass of the donor and the efficiency of mass transfer in the creation of Ba-enriched BSSs. This mass-dependent efficiency might explain why the Ba-enriched BSSs in the clusters that we have observed occupy the same region of the HR diagram, and furthermore might be applied to the distribution of all BSSs in these clusters. Future modeling and observational work is necessary in order to determine if this relationship extends more generally to BSSs.

\section{Summary}
\label{sec:summary}

We have conducted a Ba abundance survey of the BSSs of the open clusters NGC 7789 (1.6 Gyr) and M67 (4 Gyr). We find six BSSs significantly enriched in Ba compared to the MS in NGC 7789, and four Ba-enriched BSSs in M67. We take such enrichment to be indicative of AGB mass transfer.

In order to study the frequency of Ba enrichment and the properties of enriched systems as a function of cluster age, we combine our observational results with previous Ba abundance studies of the BSSs of the open clusters NGC 6819 (2.5 Gyr; \citetalias{Milliman2015}) and NGC 188 (7 Gyr; \citealt{Milliman2016}). We find in the three youngest of these open clusters an anticorrelation in the degree of Ba enrichment with cluster age, consistent with reduced barium production in AGB stars of decreasing mass. The absence of Ba enrichment in the old open cluster NGC 188 also is consistent with predictions that solar-metallicity AGB stars of less than 1.3 \Msolar\, do not undergo third dredge-up and consequently do not produce Ba enrichment. 

We also find that 40$\pm$16\% of the Ba-enriched BSSs are in binary systems with $P_{\mathrm{orb}}<5000$~days. The long orbital periods of most of these systems are indicative of mass transfer from AGB companions in binary systems. The non-velocity-variable Ba-enriched BSSs may have been created through other means of mass transfer, such as wind mass transfer or WRLOF. Theory predicts that these forms of mass transfer can result in orbital periods much longer than our RV detection limit.

We also compare our observed Ba-enriched BSSs to previous studies of field Ba dwarfs (\citealt{Escorza2017, Escorza2019}). We find that our Ba-enriched BSSs are systematically hotter and more luminous than the field Ba dwarf population, indicative of higher masses in the BSSs. We also find that the isochronal masses of the Ba-enriched BSSs are consistent with the progenitor masses of evolved Ba giants, which suggests a common origin through AGB mass transfer.

 The location of the Ba-enriched BSSs on the HR diagram is strikingly similar, regardless of the cluster age or the distance from the MSTO of the host cluster. The consistent location of the Ba-enriched BSSs implies a connection between the mass of the donor star and the efficiency of mass transfer, in that less massive AGB stars undergo more efficient mass transfer and produce BSSs more widely separated from the MSTO of their host clusters. Future work is needed in order to probe this connection in detail.
 
 Given the preponderance of long orbital periods ($P_{\mathrm{orb}}\sim1000$~days) among the BSSs of M67 and NGC 188, as well as the frequency of Ba enrichment among the BSSs of NGC 7789, NGC 6819, and M67, it may be the case that AGB mass transfer is the dominant mechanism of BSS formation in open clusters older than about 1 Gyr, even among non-velocity-variable stars. Similar observations of BSSs in more open clusters are needed to probe this mechanism further.
 
With this work we have a first look at AGB mass transfer and its role in the creation of BSSs over a span of $\sim$10 Gyr. Future detailed modeling work and comprehensive abundance studies, particularly of other mass transfer tracers such as carbon, nitrogen, and lithium are needed in order to create a complete picture of mass transfer in open clusters.

\vspace{1 cm}

The University of Wisconsin-Madison authors acknowledge funding support from NSF AST-1714506 and the Wisconsin Alumni Research Foundation. ACN acknowledges funding support from Sigma Xi grant G20201001111059492. The authors also gratefully acknowledge the many Wisconsin undergraduate and graduate students who have contributed to the WIYN Open Cluster Study radial-velocity database. The authors would also like to thank Dr.~Ana~Escorza and Dr.~Onno~Pols for the enriching conversations which have strengthened this manuscript.

This work has made use of data from the European Space Agency (ESA) mission {\it Gaia} (\url{https://www.cosmos.esa.int/gaia}), processed by the {\it Gaia} Data Processing and Analysis Consortium (DPAC, \url{https://www.cosmos.esa.int/web/gaia/dpac/consortium}). Funding for the DPAC has been provided by national institutions, in particular the institutions participating in the {\it Gaia} Multilateral Agreement.

This work has made use of Astropy (\url{http://www.astropy.org}), a community-developed core Python package for Astronomy \citep{astropy2013, astropy2018, astropy2022}. This work has also made use of PyAstronomy (\citealt{Czesla2019}).

This work was conducted at the University of Wisconsin-Madison which is located on occupied ancestral land of the Ho-chunk Nation, and observations for this work were conducted on the traditional lands of the Tohono O'odham Nation. We respect the inherent sovereignty of these nations, along with the other 11 First Nations in Wisconsin as well as the southwestern tribes of Arizona. We honor with gratitude these lands and the peoples who have stewarded them, and who continue to steward them, throughout the generations.

\vspace{1cm}
\bibliography{abun_refs}{}
\bibliographystyle{aasjournal}

\end{document}